\documentclass[traditabstract, twocolumn]{aa}

\usepackage{graphicx}
\usepackage[T1]{fontenc}
\usepackage{ae,aecompl}
\usepackage{lscape}
\usepackage{txfonts}
\usepackage{natbib}
\bibpunct{(}{)}{;}{a}{,}{,}
\begin{document}

\title{Measurements with STEREO/COR1 data of drag forces \\ acting on small-scale blobs falling in the intermediate corona}

\author{S. Dolei\inst{1}, A. Bemporad\inst{2} \and D. Spadaro\inst{1}}

\institute{INAF -- Catania Astrophysical Observatory, Catania, Italy\\
           \email{sdo@oact.inaf.if}
           \and
           INAF -- Turin Astrophysical Observatory, Pino Torinese (TO), Italy\\
           \email{bemporad@oato.inaf.it}}
   \date{Received 4 January 2013 / Accepted 20 December 2013}

\abstract{In this work we study the kinematics of three small-scale (0.01~R$_\odot$) blobs of chromospheric plasma falling back to the Sun after the huge eruptive event of June 7, 2011. From a study of 3-D trajectories of blobs made with the Solar TErrestrial RElations Observatory (STEREO) data, we demonstrate the existence of a significant drag force acting on the blobs and calculate two drag coefficients, in the radial and tangential directions. The resulting drag coefficients $C_D$ are between 0 and 5, comparable in the two directions, making the drag force only a factor of 0.45~--~0.75 smaller than the gravitational force. To obtain a correct determination of electron densities in the blobs, we also demonstrate how, by combining measurements of total and polarized brightness, the H$\alpha$ contribution to the white-light emission observed by the COR1 telescopes can be estimated. This component is significant for chromospheric plasma, being between 95 and 98~\% of the total white-light emission. Moreover, we demonstrate that the COR1 data can be employed even to estimate the H$\alpha$ polarized component, which turns out to be in the order of a few percent of H$\alpha$ total emission from the blobs. If the drag forces acting on small-scale blobs reported here are similar to those that play a role during the CME propagation, our results suggest that the magnetic drag should be considered even in the CME initiation modelling.}


\authorrunning{Dolei, Bemporad and Spadaro}
\titlerunning{Measurements of drag forces in the intermediate corona}
\maketitle

\section{Introduction}
It is well known from fluid dynamics that, when a rigid body moves relative to a fluid at high Reynolds numbers, vortices form in the trailing edge of the body leading to the formation of a turbulent flow past the object and to vortex shedding. Formation of vortices corresponds to a net transfer of energy and momentum from the body to the fluid, resulting in an effective deceleration of the body itself, which is usually described by introducing an effective drag force. A similar approach has been extensively applied in solar physics to study the interaction of magnetic fluxtubes with the surrounding plasma \citep[e.g.][]{b2a}, in that the collective behaviour of fluid particles in a collisional fluid is replaced by the presence of a magnetic field in a collisionless plasma. In the solar plasma, the magnetic Reynolds number $R_m$, representing the relative importance of plasma advection over the Ohmic diffusion, is usually $R_m\gg1$ (in particular in the solar corona where $R_m\approx10^{13}$), which leads to the well-known freezing-in of the magnetic field by the highly conducting solar plasma, and probably makes the plasma turbulence ubiquitous. Historically, the analytic drag-force approach has been adopted to study the dynamics of solar plasmas in many different environments, like the formation of sunspots \citep[e.g.][]{parker1979}, the motion of flux tubes in the convection zone \citep[e.g.][]{parker1979}, the motion of loops in the corona \citep[e.g.][]{cargill1994}, and the acceleration and propagation of magnetic clouds \citep[e.g.][]{chengarren1993}.

More recently, the same approach has been applied by \citet[][]{toriumi2011} to study the emergence of fluxtubes from the photosphere, by \citet[][]{b7} to study the downflows in quiescent prominences, and by \citet[][]{chenschuck2007} to study the damping of loop oscillations in the corona. Moreover, over the last few decades many authors applied the concept of a magnetic drag force to study the interplanetary propagation of solar eruptions (or coronal mass ejections -- CMEs); CMEs can propagate in the intermediate corona (which extends from 1.5 to 5 solar radii) at velocities of up to 2500~km~s$^{-1}$ \citep{b16,b5}, while at 1 AU velocities tend to be closer to that of the solar wind \citep[SW;][]{b6}, around 500~--~800~km~s$^{-1}$. The reason for this deceleration is that during their propagation the initial acceleration ceases and the CME motion becomes dominated by the interaction with the solar wind. \citet{b2a} proved that as the flux tube moves through the external plasma, its shape becomes distorted and reconnection can take place between the flux tube and external fields. As recently pointed out by \citet[][]{matthaeusvelli2011}, ``the problem of the interaction of CMEs with background interplanetary medium is analogous to high Reynolds number turbulent flow past an obstacle, but more complex because of expected kinetic effects.'' The coupling occurs when there is a unidirectional external field component in the direction of flux tube propagation and drag coefficients ($C_D$) parameterize this interaction. These phenomena may significantly modify the interplanetary propagation of CMEs, thus affecting the expected arrival time of solar storms on Earth, with very important consequences for space weather predictions.

The above discussion shows that the study of magnetic drag forces acting on plasma elements is an interdisciplinary topic, having a broad range of possible applications for solar physics plasmas and beyond. Nevertheless, despite the potential impact, for instance, in the study of the early propagation phase of CMEs, measurements of drag coefficients in the intermediate corona are very rare. The kinematics of plasma blobs propagating in the corona along coronal streamers were already extensively studied with SOHO/LASCO data \citep[see][]{sheeley1997,wang2000} and these blobs are believed to form because of magnetic reconnection occurring along the streamer current sheets. A very interesting class of events was also reported by \citet[][]{wangsheeley2002}, who discovered material within the bright core of a CME that collapse towards the Sun; the authors interpreted this material as plasma that was gravitationally and/or magnetically bound to the Sun. The authors assumed different values of the magnetic drag coefficient to show that drag forces can account for the asymmetry between the ascending and descending portions of the observed fall-back trajectories, but no measurements of $C_D$ were provided.

In this work, we study the 3-D trajectories of three small-scale (0.01~R$_\odot$) chromospheric plasma blobs observed in the STEREO/COR1 field of view after the huge solar eruption of June 7, 2011. The trajectories and kinematics of these blobs are reconstructed via a triangulation technique \citep{b9}. These measurements are then combined with electron density and mass estimates to investigate the effects of drag forces on the blobs. Section \ref{observations} provides a description of the observations and of the technique we employed to reconstruct the 3-D trajectories of the blobs. In Sect. \ref{analysis}, we present the analysis procedure and treat in detail the calculation of two drag coefficients, starting from the estimate of the electron density in the blobs. In addition, we focus our attention on the emitting processes and, through the degree of polarization, prove that a significant fraction of the total emission is due to H$\alpha$, in both polarized and unpolarized components, mixed with emission due to Thomson scattering of photospheric radiation by free electrons inside the blob. Discussions of our results and final conclusions are given in Sect. \ref{conclusions}.

\section{Observations}
\label{observations}
\begin{figure}
\centering
\includegraphics[bb=15 0 425 330, clip, width=0.4\textwidth]{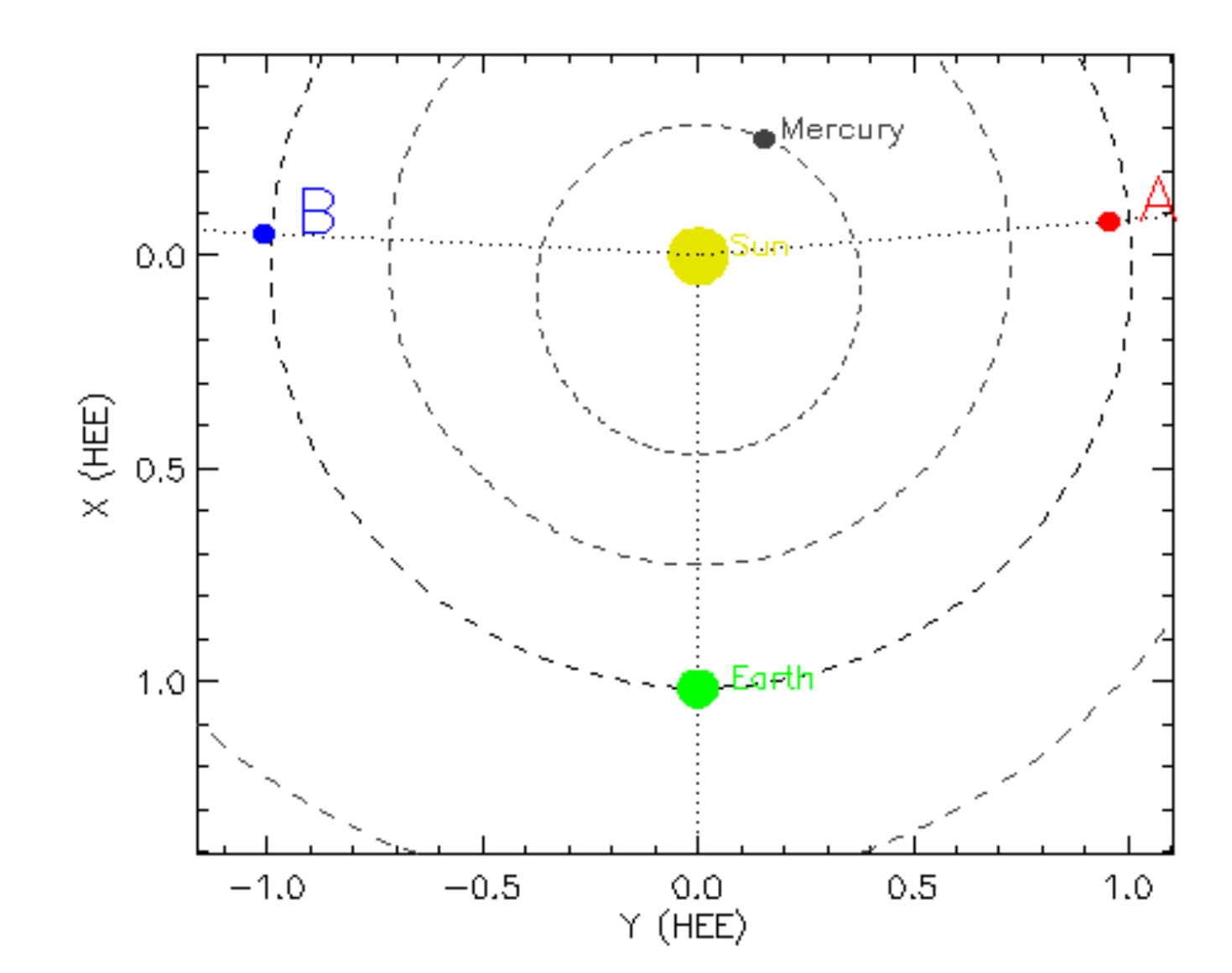}
\caption{Positions of the STEREO Ahead (red) and Behind (blue) spacecraft relative to the Sun (yellow) and the Earth (green) on June 7, 2011. The dotted lines show the angular displacements of the planets from the Sun. Distances are in Astronomical Units in a cartesian heliocentric Earth ecliptic (HEE) coordinate system.}
\label{STEREO}
\end{figure}

Coronal mass ejections are usually observed in white-light images provided by coronagraphs, such as the {\em Large Angle and Spectrometric Coronagraph} (LASCO) \citep{b1} onboard the {\em Solar and Heliospheric Observatory} (SOHO). The coronagraphs provide us with a view of a CME projected on the plane of the sky (POS). From the LASCO images, we are not able to infer the 3-D structure of a CME. The new generation data from the {\em Solar TErrestrial RElations Observatory} (STEREO) \citep{b11}, which was launched in October 2006, provide us with the first ever stereoscopic images of the Sun's atmosphere. The two STEREO spacecraft, Ahead and Behind, orbit the Sun at approximately 1 AU near the ecliptic plane with a separation angle between them increasing at a rate of about 45$^{\circ}$/year. Figure \ref{STEREO} plots the positions of the STEREO Ahead (red) and Behind (blue) spacecraft at observation date relative to the Sun (yellow) and Earth (green). In particular, on June 7, 2011 the angle between the two STEREO spacecraft was about 172$^{\circ}$. The COR1 coronagraph aboard both STEREO spacecraft, whose field of view goes from 1.4 to 4.0~R$_\odot$, takes simultaneous sequences of three polarized images that can be combined to give total brightness or polarized brightness images. From these images, the spatial location of each selected blob is derived via triangulation (also known as the tie pointing technique). One can also use observations from the C2 coronagraph aboard LASCO to perform triangulation with STEREO images. The LASCO position, along the Sun-Earth line, provides an additional constraint for stereoscopic studies.

\subsection{Selected sample}
\begin{figure*}
\centering
\includegraphics[width=0.3\textwidth]{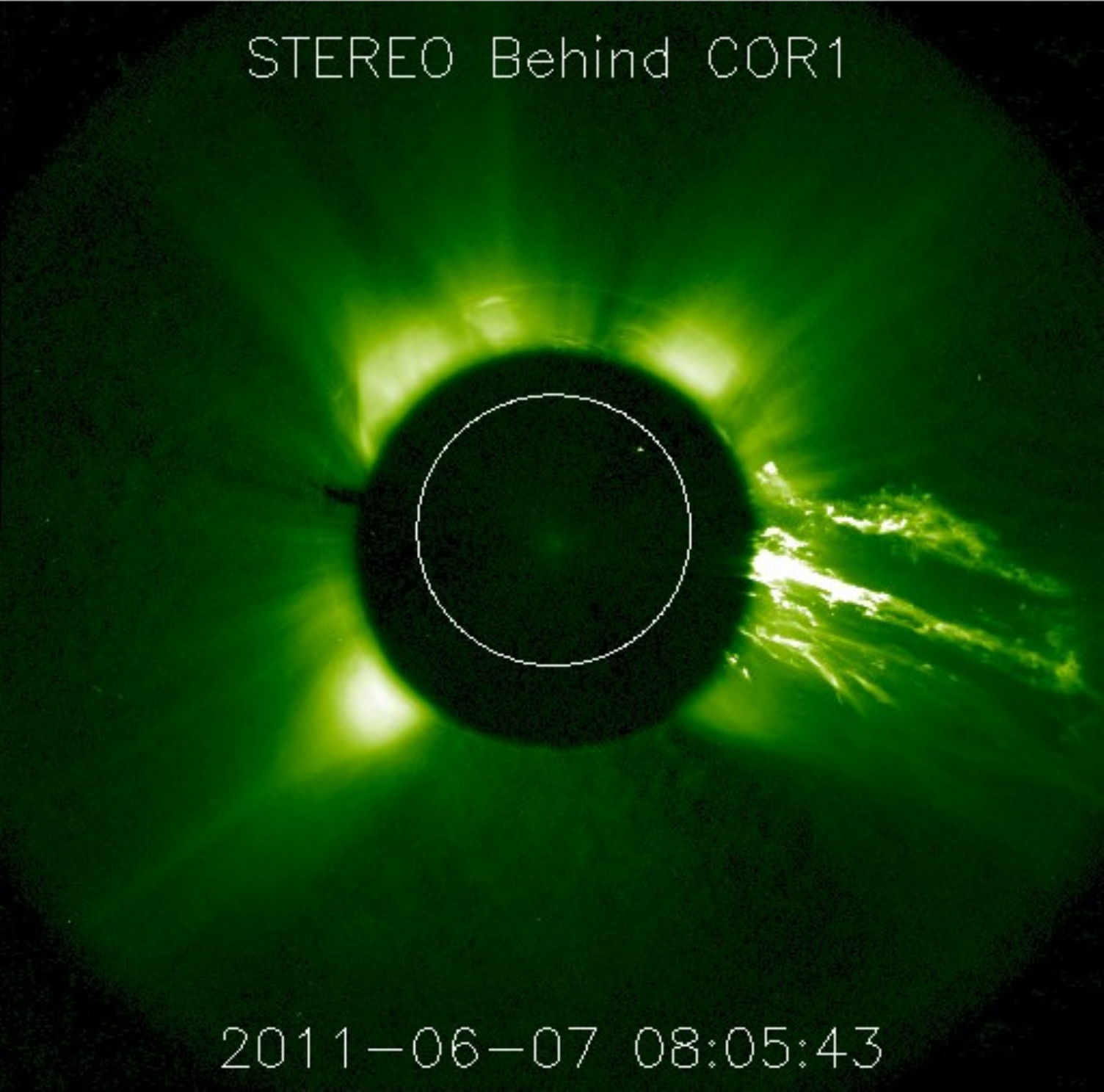}
\qquad
\includegraphics[width=0.3\textwidth]{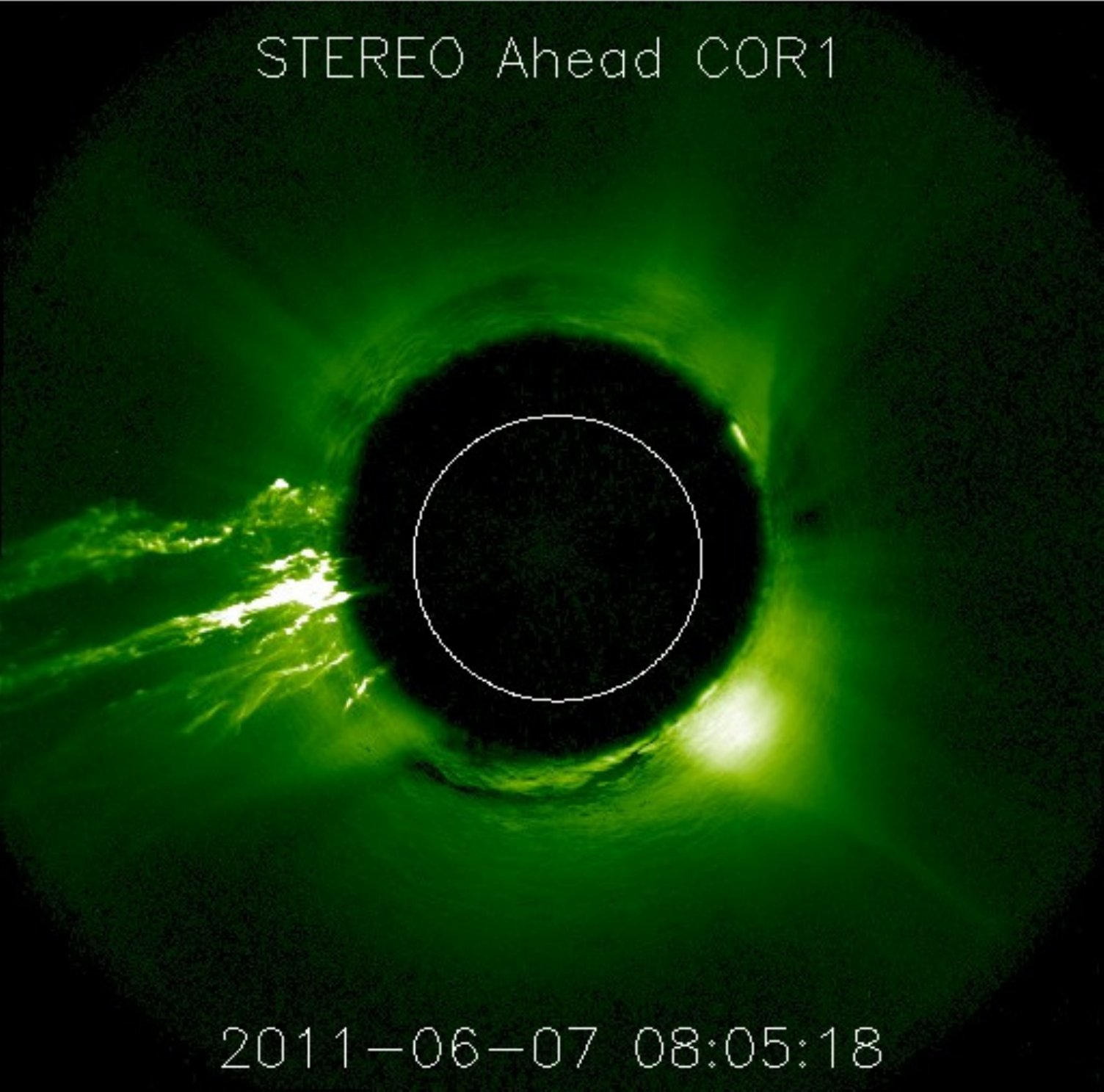}
\qquad
\includegraphics[width=0.3\textwidth]{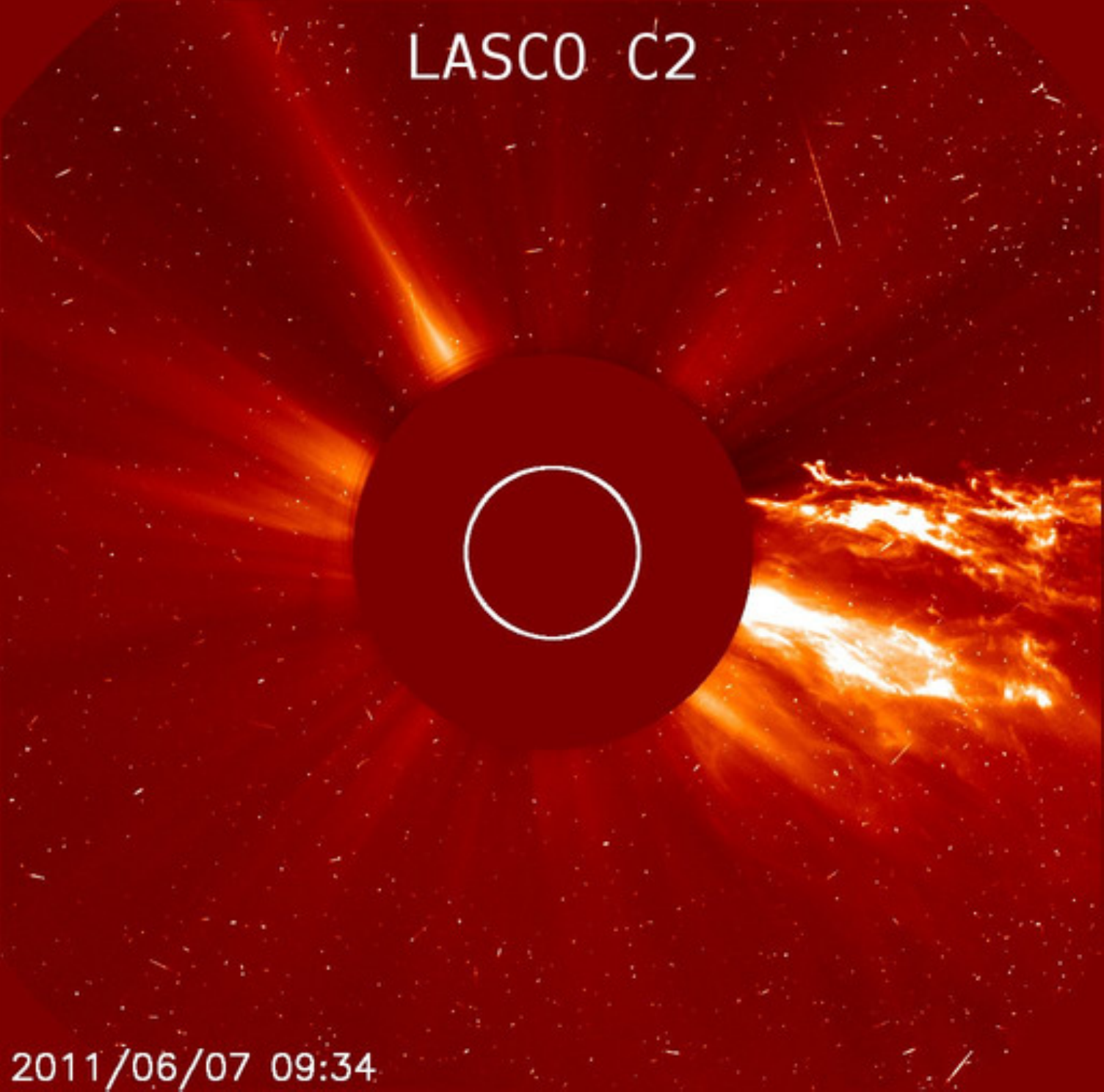}
\caption{Images acquired at 8:05 UT by STEREO/COR1-B (left) and STEREO/COR1-A (middle), and at 9:34 UT by LASCO/C2 (right), showing the huge amount of chromospheric plasma ejected during the solar eruption of June 7, 2011. In particular, in the C2 frame, two out of the three blobs studied in this work are sketched at a position angle of about 290 degrees. The white circle outlines the limb of the solar disc.}
\label{views}
\end{figure*}

\begin{figure*}
\centering
\includegraphics[width=0.3\textwidth]{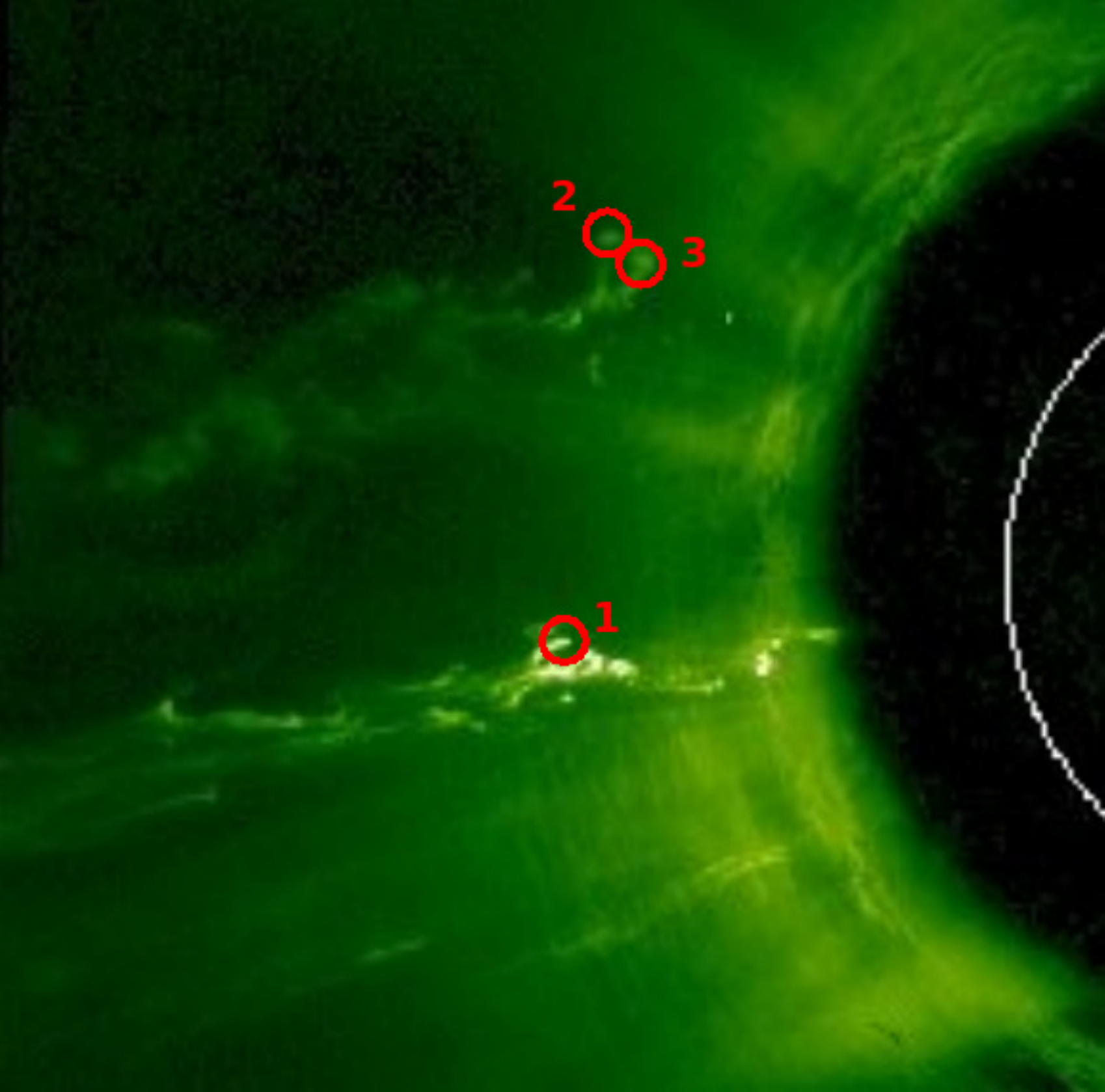}
\qquad
\includegraphics[width=0.3\textwidth]{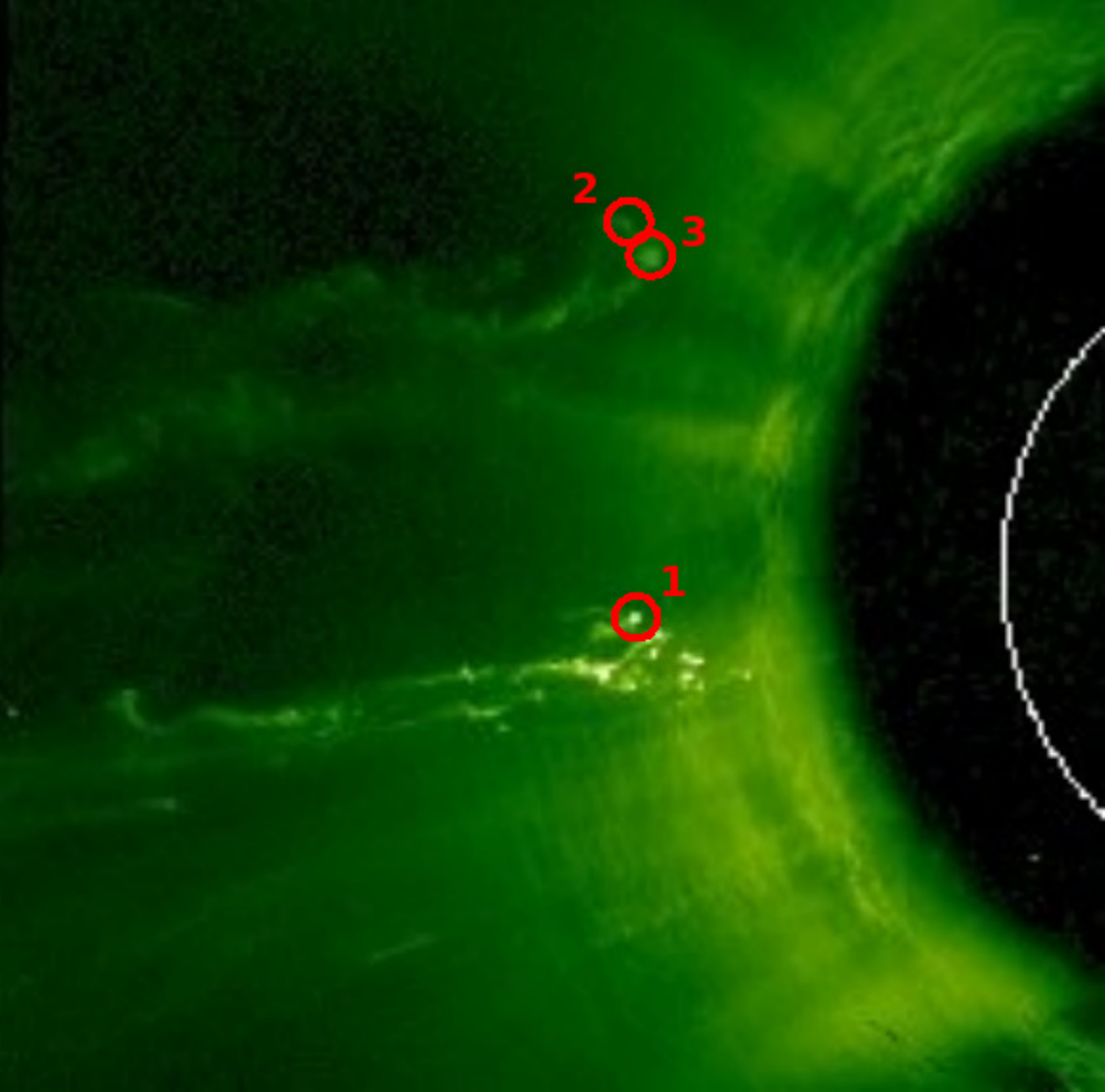}
\qquad
\includegraphics[width=0.3\textwidth]{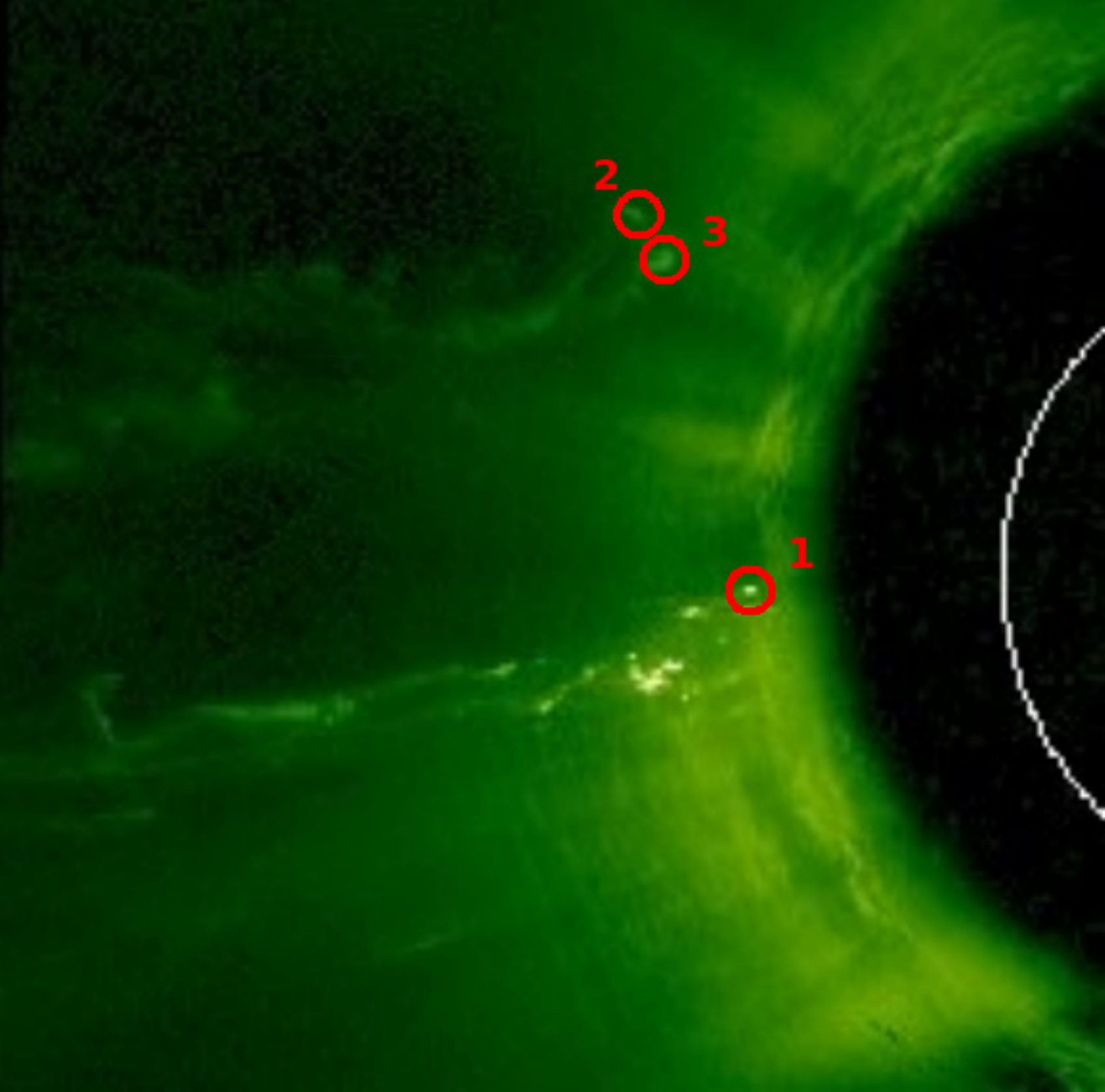}
\caption{Zoom-in on the STEREO/COR1-A images acquired at 9:35 UT (left), 9:55 UT (middle), and 10:15 UT (right). Red circles show the location of the three falling blobs employed for this study at the three different times. The white line outlines part of the limb of the solar disc.}
\label{zoom}
\end{figure*}

\begin{figure*}
\centering
\includegraphics[width=0.45\textwidth]{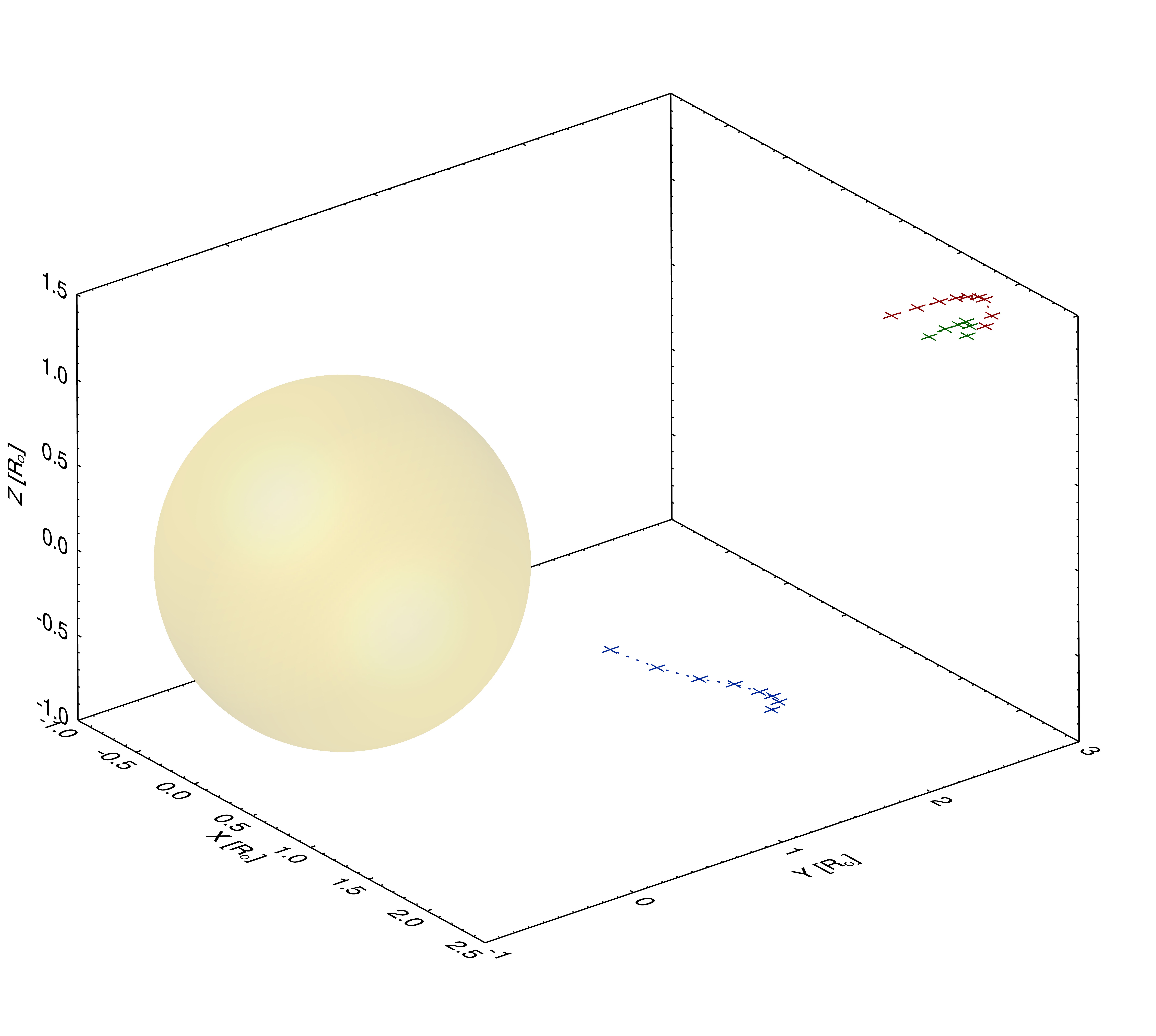}
\qquad
\includegraphics[width=0.45\textwidth]{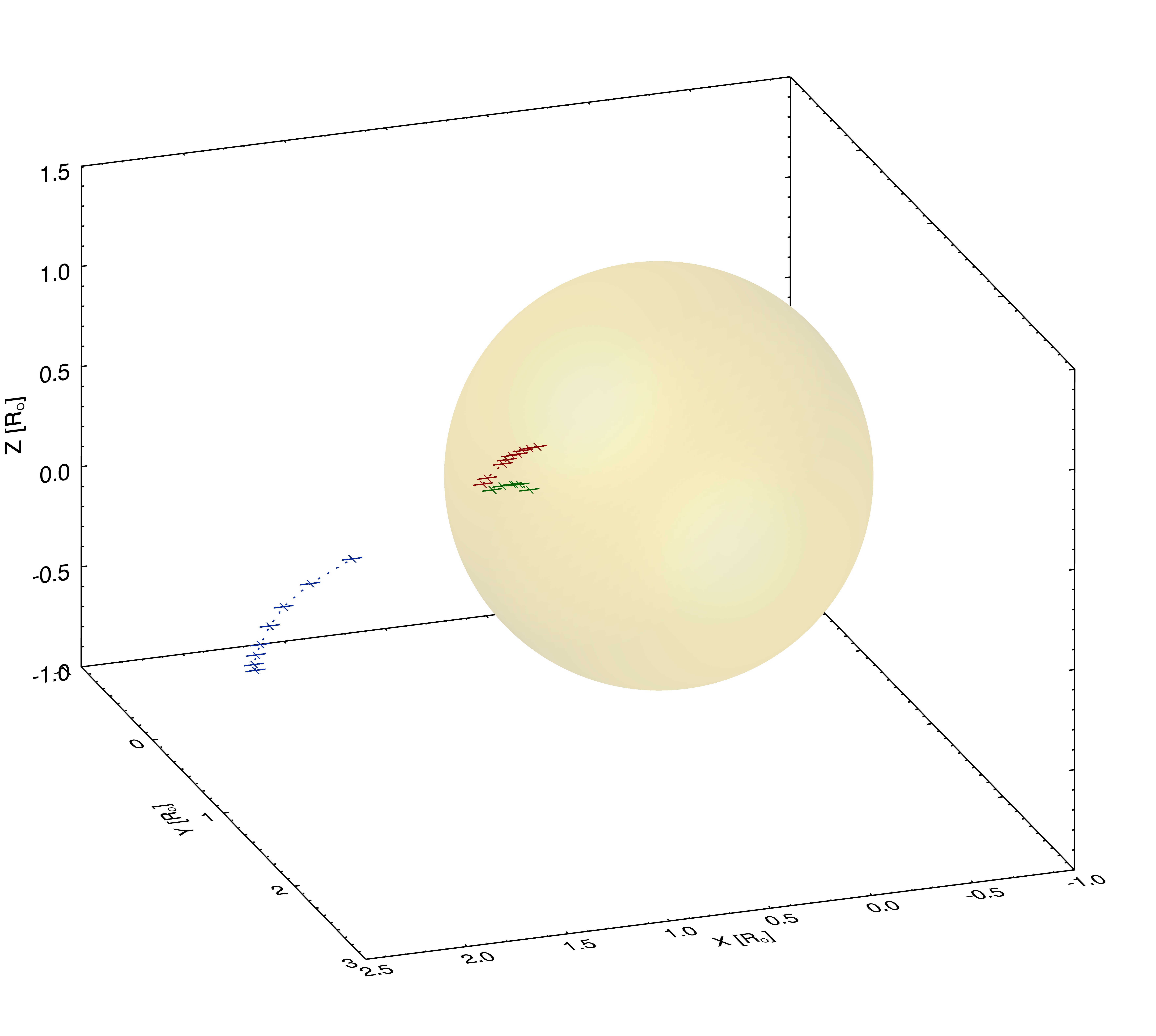}
\caption{Two different viewpoints of the falling 3-D trajectories derived via triangulation for blob 1 (blue), blob 2 (red), and blob 3 (green). The reference system, centred on the Sun, has the $x$-axis pointing towards the Earth, and the $y$- and $z$-axes pointing towards the solar east and solar north, respectively, as seen from the Earth.}
\label{3d}
\end{figure*}

Figure \ref{views} shows three snapshots of the eruption on June 7, 2011 captured by both the COR1-A and -B coronagraphs at 08:05 UT and by the C2 coronagraph at 9:34 UT. These images show a large amount of irregularly distributed plasma ejected and partially falling back to the Sun. In this work, we limit our analysis to some of the brightest blobs (those that are three orders of magnitude brighter than the quiescent corona) in order to have a better determination of the location in each frame of the blob's centroid. In particular, by looking at STEREO/COR1 and SOHO/LASCO-C2 images acquired between 06:00 UT and 14:00 UT, with a 5 minutes time cadence for COR1 (95 frames for both spacecraft) and a 15 minutes time cadence for LASCO-C2 (33 frames), we select three blobs that remain coherent during the fall back to the Sun. We consider them as spheres with radius of about 0.01~R$_\odot$, corresponding to the COR1 pixel size. These blobs are marked as 1, 2, and 3 in Figure \ref{zoom}, which shows a sub-field of the COR1-A images at 09:35 UT, 09:55 UT, and 10:15 UT. We use sequences with a 15 minutes time cadence, specifically eight COR1 frames from 08:30 UT to 10:15 UT for blob 1, nine COR1 frames from 09:20 UT to 11:35 UT for blob 2, and six COR1 frames from 09:20 UT to 10:50 UT for blob 3. In each frame, we identify four or five points as the most probable position of the blob centroid and triangulate them; then we average among these triangulated points to minimize the indetermination of the spatial location of the blobs. Finally, the reconstruction of the falling trajectories are made by tracking blobs through these frames. We use the routine {\sf scc\_measure.pro} available in the STEREO package of the {\em SolarSoftware} library which, after reading in a pair of STEREO/COR1-A and -B images, is able to track the line of sight (LOS) of a point selected in one image pair into the field of view of the second image. The reconstructed falling 3-D trajectories are shown in Figure \ref{3d}; plasma falls back to the Sun with an average inclination with respect to the radial direction of 29.7$^{\circ}$ for blob 1 (blue), 28.7$^{\circ}$ for blob 2 (red), and 27.0$^{\circ}$ for blob 3 (green).

\section{Analysis procedure}
\label{analysis}
As \citet{b2a} pointed out, reconnection between the CME and the external field, whereby the CME velocity approximates that of the solar wind, occurs only when the external field is in the direction of flux tube propagation. In this work, we assume that the ambient magnetic field, left open by the huge eruption on June 7, 2011, has radial direction and that blobs are propagating across this radial field. This assumption is simply dictated by the strong inclinations we find (of almost $30^\circ$) between the blob trajectories and the radial direction, in the altitude range 2.0~--~3.7~R$_\odot$. The magnetic field of the solar corona is not expected to have a significant tangential component at these altitudes: for instance, potential field reconstructions assume that above a spherical source surface all field lines are radial. Usually, a good agreement between extrapolations and the distribution of coronal white-light structures is found by setting a radius of this surface at 2.5~R$_\odot$ \citep[see e.g.][and references therein]{saez2007}.

\begin{figure}
\centering
\includegraphics[bb=50 20 350 450, clip, width=0.3\textwidth]{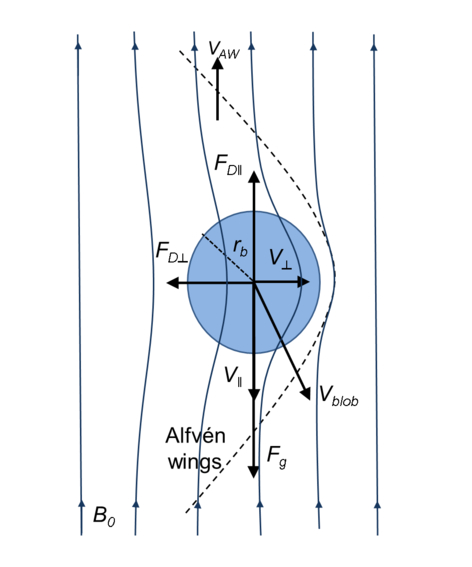}
\caption{An illustration showing the plasma blob propagating with velocity $V_{blob}$ across the post-eruption radial field $B_0$ and being subjected to the gravitational force $F_g$ and to the drag forces parallel $F_{D\parallel}$ and perpendicular $F_{D\perp}$ to the magnetic field. Propagation of Alfv\'en waves generates two wings and carries away momentum and energy.}
\label{aw}
\end{figure}

This assumption allows us to determine the radial component of the drag force $F_{D\parallel}$ (see Sect.~\ref{rad_kin}). On the other hand, the tangential component of the drag force $F_{D\perp}$ (see Sect.~\ref{tan_kin}) can be derived by assuming that blobs can be thought of as packets which deform the ambient field as they push their way through it. As recently pointed out by \citet{b7}, who provided a droplet model for plasma downflows observed in a quiescent prominence, although the magnetic tension attempts to restore the undisturbed conditions after the passage of the packet, the temporary distortion of the field lines excites Alfv\'en waves carrying energy and momentum away from the interaction volume. This loss of momentum constitutes a magnetic drag force that opposes the motion. Figure \ref{aw} shows the model we present here, which is a rearrangement of the model presented by \citet{b7}: blobs move into the solar wind and warp the radial field lines.

\begin{figure}
\centering
\includegraphics[width=0.45\textwidth]{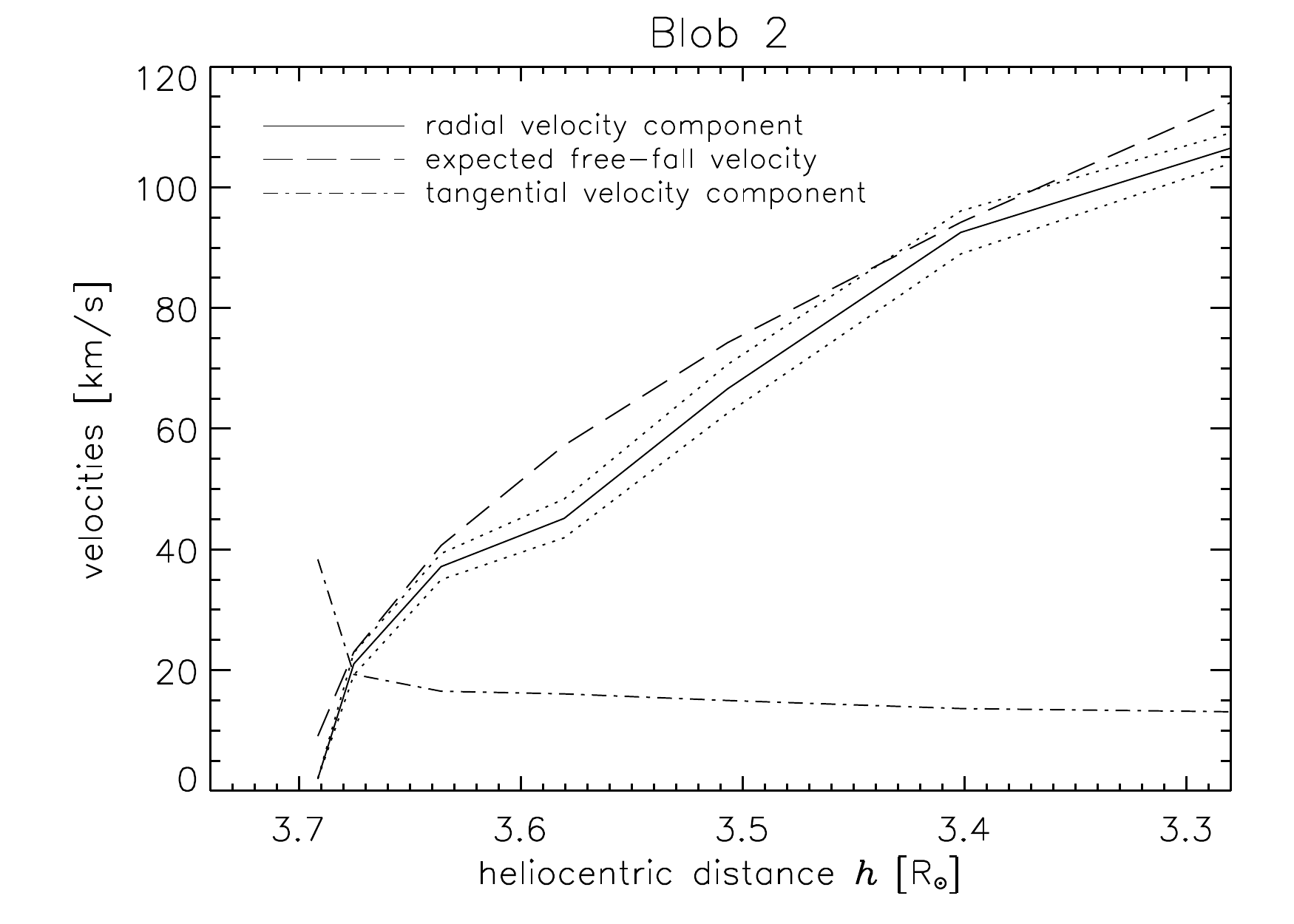}
\caption{Falling velocities along the blob 2 trajectory. The plot shows that the radial velocity component (solid line) remains smaller than the expected free-fall velocity (dashed line), even if it ranges between its upper and lower limits (dotted lines), and that the tangential velocity component (dash-dotted line) decreases at lower altitudes, and a tangential deceleration may be derived.}
\label{velocity}
\end{figure}

We estimate the radial and tangential speed components of the blobs along their falling trajectories towards the Sun. For instance, in the case of blob 2, Figure \ref{velocity} shows that the tangential velocity (dash-dotted line) decreases at lower altitudes in disagreement with the conservation of angular momentum and the radial velocity (solid line ranging between the dotted lines) is smaller than free-fall velocity (dashed line) due only to the solar gravitational field. The kinematics of the blobs are affected by the magnetic field through drag forces. We derive two drag coefficients, starting from the radial motion, due to the combined action of the gravitational and radial drag forces, and from the tangential motion, affected by the tangential drag force due to the magnetic tension produced by the bending of the field lines because the blob crosses the ambient magnetic field.

\subsection{Analysis of white-light emissions}

In the following we discuss how the white-light COR1 intensities have been employed to estimate not only the electron densities in the blobs, but also the absolute H$\alpha$ total and polarized brightness emitted by the blobs.

\subsubsection{Estimate of blob electron density}
\label{bdc}
\begin{figure}
\centering
\includegraphics[width=0.45\textwidth]{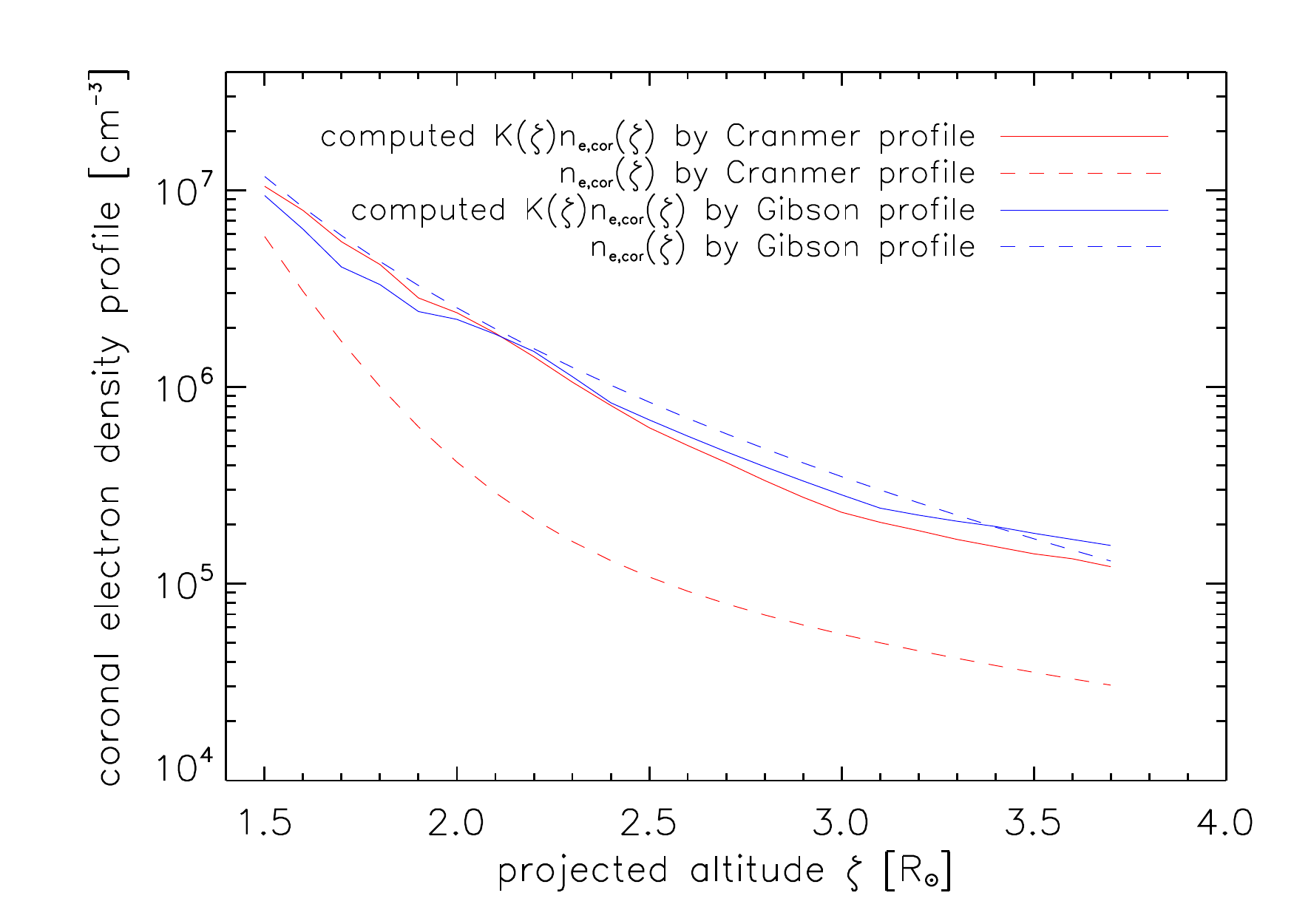}
\caption{The computed coronal electron density profiles $K(\zeta)\,n_{e,cor}(\zeta)$ (blue and red solid lines) on the plane of the sky (i.e. for $z=0$), where the $n_{e,cor}(\zeta)$ functions (blue and red dashed lines) are given in Eq. (\ref{profile}) for $h=\zeta$.}
\label{comparison}
\end{figure}

For our purposes, we need to estimate the density in the blobs, by knowing the blob location determined via triangulation, assuming a coronal electron density profile and measuring the brightness due to Thomson scattering of photospheric light by free electrons \citep{b14a,b15a,b0}. The STEREO/COR1 coronagraphs \citep{b15} observe in a 22.5 nm wide waveband centred at the H$\alpha$ line at 656.3 nm, which is the result of the electronic 3$\rightarrow$2 transition by neutral hydrogen atoms. As demonstrated by e.g. \citet{b0b}, the chromospheric material in prominence filaments associated with an erupting event is optically thin in $H\alpha$. Therefore, we assume that the $H\alpha$ radiation from the selected blobs is optically thin. Hence, in COR1 images, the measured total brightness $tB$, observed at the pixels where each blob is located, comes from a superposition along the LOS of the radiation from the three different sources: the H$\alpha$ radiation emitted by neutral H atoms in the blob $tB_{H\alpha}$, the Thomson scattering by free electrons in the blob $tB_{Th}$, and the Thomson scattering by coronal free electrons along the same LOS through the blob $tB_{Th^\prime}$. We have
\begin{equation}
tB=tB_{H\alpha}+tB_{Th}+tB_{Th^\prime}
\label{total1}
\end{equation}
and we need to remove the H$\alpha$ and coronal contributions from the COR1 intensities to estimate the real Thomson scattering total brightness and the right blob density. In each pixel through which the blob propagates, the brightness measured at a projected altitude $\zeta$ is given by a LOS integration of the inferred electron density profile multiplied by a geometrical function; this integration is then split into two integrals: one performed over a coronal length L and the other over a $2r_b$ thickness across the blob, where $r_b$ is the radius of the blob (assumed to be spherical) about equal to 0.01~R$_\odot$, as inferred from the COR1 images. The three contributions expressed in Eq. (\ref{total1}) then become
\begin{equation}
tB=tB_{H\alpha}+\int_{2r_b}B(z)\,n_{e,blob}(z)\,dz+\int_{L-2r_b}B(z)\,n_{e,cor}(z)\,dz,
\label{total2}
\end{equation}
where $B(z)$ is the total brightness for one electron and $n_{e,blob}(z)$ and $n_{e,cor}(z)$ are the density profiles along the LOS coordinate $z$, respectively, in the blob and in the corona. We use the routine {\sf eltheory.pro} available in the {\em SolarSoftware} library, which uses the Thomson electron scattering theory to compute the value of $B(z)$, once the distance of the electron from the centre of the Sun is known. If the angle brackets $\langle\,\Gamma\,\rangle_\gamma$ denote the average value of the quantity $\Gamma(z)$ along a path of length $\gamma$ along $z$, then the integrals in Eq. (\ref{total2}) become
\begin{equation}
tB\simeq tB_{H\alpha}+\langle B\,\rangle_{2r_b}\langle\,\!n_{e,blob}\,\!\rangle_{2r_b}2r_b+tB_{cor},
\label{total3}
\end{equation}
where $tB_{cor}\simeq tB_{Th^\prime}$, because of the very small LOS extension of the blob, and $n_{e,blob}=\langle\,\!n_{e,blob}\,\!\rangle_{2r_b}$ is the blob density. To estimate the real coronal density distribution we make use of two different density radial profiles $n_{e,cor}$ available in the literature, representing two different extremes of electron density values. In particular, we set as input for our calculations either the coronal hole density distribution modeled by \citet{b3} or the streamer density profile at solar minimum derived by \citet{b4a}, respectively given by
\begin{equation}
n_{e,cor}(h)=\left\{\begin{array}{l}
10^5\left(3890\,h^{-10.5}+8.69\,h^{-2.57}\right) \\
\\
10^8\left(77.1\,h^{-31.4}+0.954\,h^{-8.30}+0.550\,h^{-4.63}\right),
\end{array}\right.
\label{profile}
\end{equation}
where the heliocentric distance $h$ is in units of R$_\odot$ and the density is in cm$^{-3}$. The aim is to reproduce the COR1-A post-eruption white-light brightness, specifically at 10:30 UT along the 130$^\circ$ position angle (measured counterclockwise from the north) and averaged over 20$^\circ$. For every pixel at the projected altitude $\zeta=\sqrt{h^2-z^2}$, we assume a set of non-dimensional multiplication factors $K(\zeta)$ with values between 0.2 and 10. Then, by integrating at each projected altitude all the $K(\zeta)\,n_{e,cor}(h)$ density profiles along the LOS resulting from the assumption of all the possible $K(\zeta)$ values between 0.2 and 10, we determine the $K(\zeta)$ factor that best reproduces the COR1 observed white-light intensity. Hence, the derived $K(\zeta)$ values represent at each projected altitude the multiplication factors of the density profile (given in Eq. (\ref{profile})) required to reproduce the observed white-light intensity. With this technique we derive an array of multiplication factors $K(\zeta)$ for different projected altitudes $\zeta$: the resulting densities on the plane of the sky (i.e. for $z=0$) as a function of $h=\zeta$ are given by $K(\zeta)\,n_{e,cor}(\zeta)$. Figure \ref{comparison} shows the resulting density profiles $K(\zeta)\,n_{e,cor}(\zeta)$ (blue and red solid lines) obtained by assuming the two different functions $n_{e,cor}(\zeta)$ given in Eq. (\ref{profile}) (blue and red dashed lines). The similarity of results implies that the initial choice of the coronal density distribution has no significant effect, because the two curves are the same on average to within less than 10\%. Once the coronal density is derived, we are able to calculate the coronal total brightness contribution $tB_{cor}$; finally, by measuring $tB$ from the COR1 images, we need only to estimate the unknown quantity $tB_{H\alpha}$ to determine $n_{e,blob}$, according to Eq. (\ref{total3}). To this end, we include one more condition.

\citet{b0b} have expressed the H$\alpha$ emission in the optically thin regime $I_{H\alpha}$ and the emission by Thomson scattering $I_{Th}$ for a prominence filament associated with an erupting event and derived an electron temperature $T_e=$ 20000~K. In our case, the formulas become
\begin{eqnarray}
I_{H\alpha} &=& 1.06\times10^{-5}\,n_3\,2r_b
\label{athay1} \\
I_{Th} &=& \sigma_e\,W\,J_{tot}\,n_{e,blob}\,2r_b,
\label{athay2}
\end{eqnarray}
where $n_3$ is the population density of the third principal quantum level in the hydrogen atom, $\sigma_e=6.65\times10^{-25}$ cm$^2$ is the electron scattering cross section, $W$ is the dilution factor given by the ratio between the mean intensity of the radiation incident from the solar surface and the intensity emitted from the solar disc centre and $J_{tot}=4.88\times10^8$ ergs cm$^{-2}$ s$^{-1}$ sr$^{-1}$ is the photospheric intensity integrated over the COR1 filter passband and centred at 656.3 nm. Above 2 R$_\odot$ the limb darkening is negligible and $W$ only has geometrical properties and is equal to
\begin{equation}
W=\frac{1}{2}\left[\,1-\sqrt{1-\frac{1}{(1+h)^2}}\,\right]
\end{equation}
with the heliocentric distance $h$ in R$_\odot$. The population density $n_3$ can be expressed by using the Boltzmann-Saha statistics \citep{b14e} as
\begin{equation}
n_3=b_3\,n_{e,blob}^2\left(\frac{2\,\pi\,m_e\,k_B\,T_e}{h^2}\right)^{-3/2}\frac{g_3}{2}\,exp\left(\frac{\chi_3}{k_B\,T_e}\right),
\label{n3}
\end{equation}
where $b_3$, $g_3$ and $\chi_3$ are, respectively, the departure factor describing the non-LTE radiative-transfer, statistical weight and excitation energy of the third principal quantum level in hydrogen atom; $m_e$ is the electron mass; $k_B$ is the Boltzmann constant; and $h$ is the Planck constant. As \citet{b10} pointed out, $b_3$ is temperature-dependent and can be fitted by a parabolic function of temperature. On the other hand, the ratio between Eqs. (\ref{athay1}) and (\ref{athay2}), through Eq. (\ref{n3}), shows a weak temperature dependence for typical prominence plasma temperatures $T_e>$~10000~K, and by extrapolating the value of $b_3$ to the electron temperature provided by \citet{b0b} ($T_e=$~20000~K), we can write
\begin{equation}
\frac{I_{H\alpha}}{I_{Th}}=\frac{tB_{H\alpha}}{tB_{Th}}=1.57\times10^{-9}\,W^{-1}\,n_{e,blob}.
\label{ratio}
\end{equation}
Nevertheless, because of the expression of $n_3$, Eq. (\ref{ratio}) holds only for static prominences. Because the plasma blob is moving with respect to the solar surface, the H$\alpha$ radiation incident on the blob and emitted by the underlying photosphere-chromosphere is Doppler-shifted with respect to the atomic absorption profile. In particular, the H$\alpha$ line has its photospheric-chromospheric counterpart in the form of a Fraunhofer absorption line and this shift results in an increase of the exciting radiation, hence in a final brightening of H$\alpha$ radiation; this is known as the Doppler brightening effect (DBE) \citep[see e.g.][]{b7a}. For this reason, in what follows, we will introduce a non-dimensional correction factor $f=f({\rm v})$ (with $f > 1$), dependent on the plasma radial velocity v, so that the H$\alpha$ radiation that one should observe in the case of a quiescent plasma is given by $tB_{H\alpha}/f({\rm v})$. The correction factor $f({\rm v})$ was computed by \citet{b14d} as a function of the heights above the solar surface; the author found that $f({\rm v})$ has values typically between 1 and 4 for an inflow or outflow radial plasma velocity v between 0 and 100~km~s$^{-1}$, up to 1~R$_\odot$, and a steady trend at higher altitudes. Hence, by introducing the factor $f$, Eq. (\ref{ratio}) has to be rewritten as
\begin{equation}
\frac{tB_{H\alpha}}{f({\rm v})\,tB_{Th}}=1.57\times10^{-9}\,W^{-1}\,n_{e,blob}.
\label{ratio2}
\end{equation}
By reasonably assuming that our selected blobs carry plasma similar to that studied by \citet{b0b} in a prominence filament or by \citet{b10} in quiescent prominences, Eqs. (\ref{total3}) and (\ref{ratio2}) correspond to an algebric system of two equations in the two unknown quantities $n_{e,blob}$ and $tB_{H\alpha}$. Solving this system for $n_{e,blob}$, we have
\begin{equation}
1.57\times10^{-9}\,W^{-1}\,f({\rm v})\,\,n_{e,blob}^2+n_{e,blob}-\frac{tB-tB_{cor}}{\langle B\,\rangle_{2r_b}2r_b}=0.
\label{quadratic}
\end{equation}
Summarising, given the observational quantities $tB$ and $r_b$, calculated $\langle B\,\rangle_{2r_b}$ and $tB_{cor}$, this is a quadratic equation, whose positive solution permits us to obtain the blob electron density and, in turn, to calculate the Thomson scattering radiation $tB_{Th}$ and the total H$\alpha$ emission $tB_{H\alpha}$ through Eq. (\ref{ratio2}).

Electron densities in the three blobs have been estimated along their 3-D trajectories at heliocentric distances between 2.0 and 3.7~R$_\odot$, where we compute a coronal electron density ranging between $\sim1.2\times10^5$~cm$^{-3}$ and $\sim2.5\times10^6$~cm$^{-3}$. Results at the different blob altitudes do not show a significant variation with height and we also conclude that the electron density for each blob is steady within the error bars. Table \ref{tab_density} gives the electron density values for the three blobs, averaged along the trajectories tracked by both STEREO spacecraft, which turn out to be in the order of $10^2$~--~$10^3$ times larger than the density of the surrounding corona.

As we have verified before, the $K(\zeta)$ parameter, whose values are returned by the COR1 post-eruption total brightness, makes our calculations almost independent of the initial choice of the coronal density profile. The radiometric uncertainty of COR1 is estimated to be $\pm$10\% \citep{b15}; however, we verify that a variation by a factor of up to 2 of the $K(\zeta)$ parameter corresponds to a difference of less than 5\% in the resulting blob density.

For comparison, Table \ref{tab_density} also shows the values of $n_{e,blob}$ derived without removing the H$\alpha$ contribution, namely by the following relationship:
\begin{equation}
n_{e,blob}=\frac{tB-tB_{cor}}{\langle B\,\rangle_{2r_b}2r_b}.
\label{over}
\end{equation}
It is easily seen that not taking into account the H$\alpha$ emission from the blob implies a strong overestimation of the electron density by at least an order of magnitude. The same uncertainty has to be considered when STEREO data are employed to measure the electron density in the core of CMEs if the strong H$\alpha$ emission of the erupting prominence is not taken into account. The technique we have described in this paper has been applied for each blob to images acquired by both STEREO spacecraft: the very good agreement between values derived for the same blob from STEREO-A and STEREO-B data demonstrates that uncertainties related to the integration along the LOS are quite small.

We note here that the values we derived for the H$\alpha$ radiance emitted by the blobs are in agreement with the hypothesis we made by using Eq. (\ref{athay1}), namely that the regime of emission is optically thin. In particular, \citet{b7b} demonstrated the existence of a correlation between the strength of H$\alpha$ emission and the correspondent optical thickness $\tau_{H\alpha}$ of the emitting plasma and found that for a line integrated intensity $tB_{H\alpha}$ smaller than $\sim10^4$~ergs~cm$^{-2}$~s$^{-1}$~sr$^{-1}$ is $\tau_{H\alpha}<10^{-1}$. In particular, in our case we infer that $tB_{H\alpha}\sim10^2$~--~$10^3$~ergs~cm$^{-2}$~s$^{-1}$~sr$^{-1}$ (see last column in Table \ref{tab_density}), hence $\tau_{H\alpha}\sim10^{-3}$~--$10^{-2}$, indicating that the plasma in the blobs is optically thin to the H$\alpha$ radiation.

\begin{table}
\centering
\renewcommand{\baselinestretch}{1.2}
\caption{Electron density of the three blobs. The values in Col. 1 (by Eq. (\ref{over})) do not include the H$\alpha$ contribution; the values in Col. 2 (by Eq. (\ref{quadratic})) include this contribution. The last column shows an estimate of the H$\alpha$ total brightness.}
\textsc{\footnotesize{
\begin{tabular}{lccc}
\hline
\multicolumn{4}{c}{Blob 1}  \\
STEREO & $n_{e,blob}$ {\rm Eq. (\ref{quadratic})} & $n_{e,blob}$ {\rm Eq. (\ref{over})} & $\log$($tB_{H\alpha}$)  \\
       & ${\rm 10}^7\,{\rm cm}^{-3}$  & ${\rm 10}^7\,{\rm cm}^{-3}$ & {\rm ergs cm$^{-2}$ s$^{-1}$ sr$^{-1}$}  \\
\hline
ahead  & 8.64 $\pm$ 0.92 & 148.4 $\pm$  7.9 & 3.159  \\
\hline
behind & 7.88 $\pm$ 0.85 & 128.1 $\pm$  7.1 & 3.141  \\
\hline
\hline
\multicolumn{4}{c}{Blob 2}  \\
STEREO & $n_{e,blob}$ {\rm Eq. (\ref{quadratic})} & $n_{e,blob}$ {\rm Eq. (\ref{over})} & $\log$($tB_{H\alpha}$)  \\
       & ${\rm 10}^7\,{\rm cm}^{-3}$  & ${\rm 10}^7\,{\rm cm}^{-3}$ & {\rm ergs cm$^{-2}$ s$^{-1}$ sr$^{-1}$}  \\
\hline
ahead  & 5.42 $\pm$ 0.54 &  84.5 $\pm$  4.1 & 2.419  \\
\hline
behind & 4.72 $\pm$ 0.81 &  55.6 $\pm$  6.6 & 2.362  \\
\hline
\hline
\multicolumn{4}{c}{Blob 3}  \\
STEREO & $n_{e,blob}$ {\rm Eq. (\ref{quadratic})} & $n_{e,blob}$ {\rm Eq. (\ref{over})} & $\log$($tB_{H\alpha}$)  \\
       & ${\rm 10}^7\,{\rm cm}^{-3}$  & ${\rm 10}^7\,{\rm cm}^{-3}$ & {\rm ergs cm$^{-2}$ s$^{-1}$ sr$^{-1}$}  \\
\hline
ahead  & 5.53 $\pm$ 0.73 &  78.2 $\pm$  5.0 & 2.402  \\
\hline
behind & 9.2  $\pm$ 1.6  & 216.3 $\pm$ 22.3 & 2.685  \\
\hline
\end{tabular}
}}
\label{tab_density}
\end{table}

\subsubsection{Estimate of H$\alpha$ polarized emission from the blobs}

In the analysis described so far we employed the total brightness observed by STEREO/COR1 instruments. Nevertheless, more information on the H$\alpha$ emission can be derived by analysing the polarized white-light component $pB$ as well. In particular, in what follows, we will show how $pB$ and $tB$ measurements from STEREO can be combined to estimate the percentage of polarization in the H$\alpha$ emission. The anisotropy of the incident photospheric light causes the observed Thomson-scattered radiation to be strongly polarized in the direction parallel to the solar limb. The presence of additional H$\alpha$ emission due to neutral hydrogen atoms in the blobs can result in a reduction of polarization. For instance, \citet{b14b} concluded that a reduction in polarization of the white-light coronal emission occurs where the raised neutral hydrogen is well embedded in the CME cloud and when it is located close to the plane of the sky of the observer.

As also pointed out by \citet{b13}, an equation similar to (\ref{total1}) can be applied when dealing with the polarized components of the white-light, since the polarized emission in COR1 images, integrated along a LOS through the optically thin chromospheric plasma ejected into the corona, is due to the same three contributions. Hence we have that
\begin{equation}
pB=pB_{H\alpha}+pB_{Th}+pB_{Th^{\prime}}
\end{equation}
and, in a similar manner to the calculation of the Thomson scattering total brightness $tB_{Th}$, we can calculate the polarized brightness $pB_{Th}$ as:
\begin{equation}
pB_{Th}=\langle p\,\rangle_{2r_b}\,n_{e,blob}\,2r_b,
\end{equation}
where $\langle p\,\rangle_{2r_b}$ is the polarized brightness for one electron averaged over the blob thickness, whose value we compute by using the method previously described for $B(z)$.

The \citet{b13} approach was to combine results from the tie pointing and Polarization Ratio \citep{b14f} techniques in order to justify the inconsistent position of some plasma ``horns'' in the 3-D reconstruction of the event they studied; they ascribed this anomaly to the presence of the low polarized H$\alpha$ component.

In this work, conversely, given the quantities $pB$ and $r_b$ directly measured from STEREO/COR1 images, we start from the determination of the blob location via triangulation, then derive the blob density and calculate $pB_{Th}$ and $pB_{Th^{\prime}} \simeq pB_{cor}$; finally, by taking into account the correction factor $f({\rm v})$ for moving prominences again, we obtain an estimate of the polarized component of H$\alpha$ emission $pB_{H\alpha}$. In addition, the total brightness calculated before allows us to measure the degree of polarization \%$pol=pB/tB$. In particular, Table \ref{tab_pol} shows, from left to right, the percentage of polarization of Thomson-scattered radiation, total radiation, and H$\alpha$ emission from the three blobs. The values are averaged along the trajectories tracked by both STEREO spacecraft. It is evident a reduction in polarization of the total radiation compared to that of the Thomson scattering, owing to the emission by neutral hydrogen atoms, whose percentage of polarization is quite low. Table \ref{tab_pol} also shows the percentage contribution of the H$\alpha$ radiation to the total brightness emission, given by \%H$\alpha=tB_{H\alpha}/(tB_{Th}+tB_{H\alpha})$. Results are in good agreement for instance with \citet{b17} and \citet{b11a}, who measured the linear polarization of H$\alpha$ radiation from prominences and found it varies within a few percent. Moreover, by the examination of a CME on August 31, 2007, \citet{b13} obtained a H$\alpha$ contribution to the total brightness emission more than 88\% with a very low polarization, less than 5\%.

\begin{table}
\centering
\renewcommand{\baselinestretch}{1.2}
\caption{Percentage of polarization of Thomson-scattered radiation, total radiation, H$\alpha$ emission and percentage of contribution of the H$\alpha$ emission to the total radiation from the three blobs.}
\textsc{\footnotesize{
\begin{tabular}{lcccc}
\hline
\multicolumn{5}{c}{Blob 1}  \\
STEREO & \%$pol_{Th}$ & \%$pol_{blob}$ & \%$pol_{H\alpha}$ & \%{\rm H}$\alpha$  \\
\hline
ahead  & 73.89 & 10.28 & 6.21 & 97.43  \\
\hline
behind & 61.07 &  8.91 & 5.08 & 97.10  \\
\hline
\hline
\multicolumn{5}{c}{Blob 2}  \\
STEREO & \%$pol_{Th}$ & \%$pol_{blob}$ & \%$pol_{H\alpha}$ & \%{\rm H}$\alpha$  \\
\hline
ahead  & 23.04 &  6.35 & 5.20 & 96.70  \\
\hline
behind & 14.23 &  4.46 & 3.83 & 95.02  \\
\hline
\hline
\multicolumn{5}{c}{Blob 3}  \\
STEREO & \%$pol_{Th}$ & \%$pol_{blob}$ & \%$pol_{H\alpha}$ & \%{\rm H}$\alpha$  \\
\hline
ahead  & 22.35 &  3.13 & 1.59 & 96.02  \\
\hline
behind & 13.47 &  3.70 & 3.27 & 97.44  \\
\hline
\end{tabular}
}}
\label{tab_pol}
\end{table}

\subsection{Analysis of 3-D kinematics}

Once the average electron densities in the blobs (hence, given their radius, the blob masses) are known, it is possible to couple this information with the accelerations measured via triangulation, thus providing an estimate for the effective forces accelerating the blobs in the corona. To this end, in what follows we will separate, for obvious reasons, the two motions projected over the radial and the tangential directions, providing an estimate for the radial and tangential drag forces acting on the blobs.

\subsubsection{Radial kinematics: magnetic drag force}
\label{rad_kin}
\begin{figure*}
\centering
\includegraphics[width=0.33\textwidth]{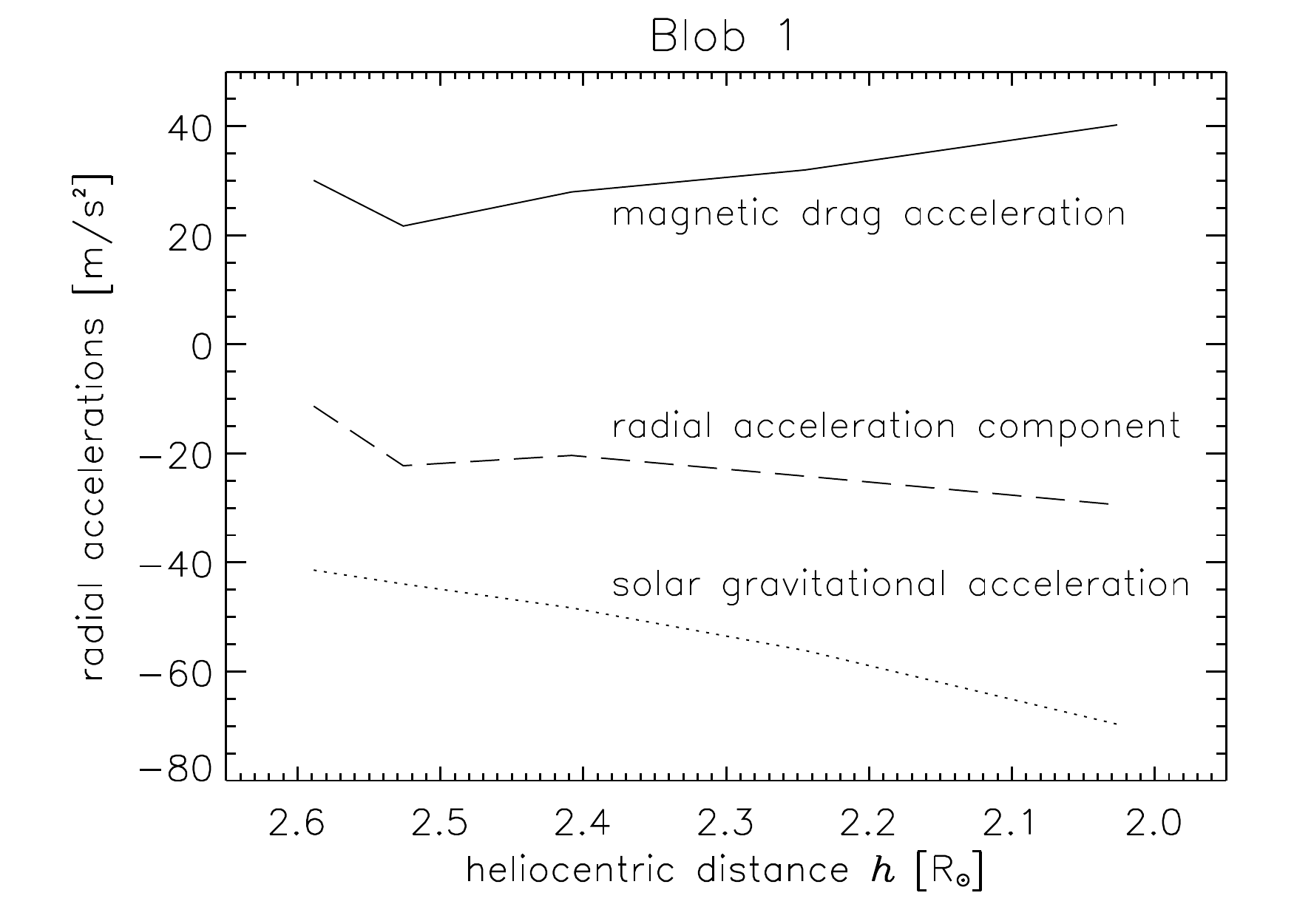}
\includegraphics[width=0.33\textwidth]{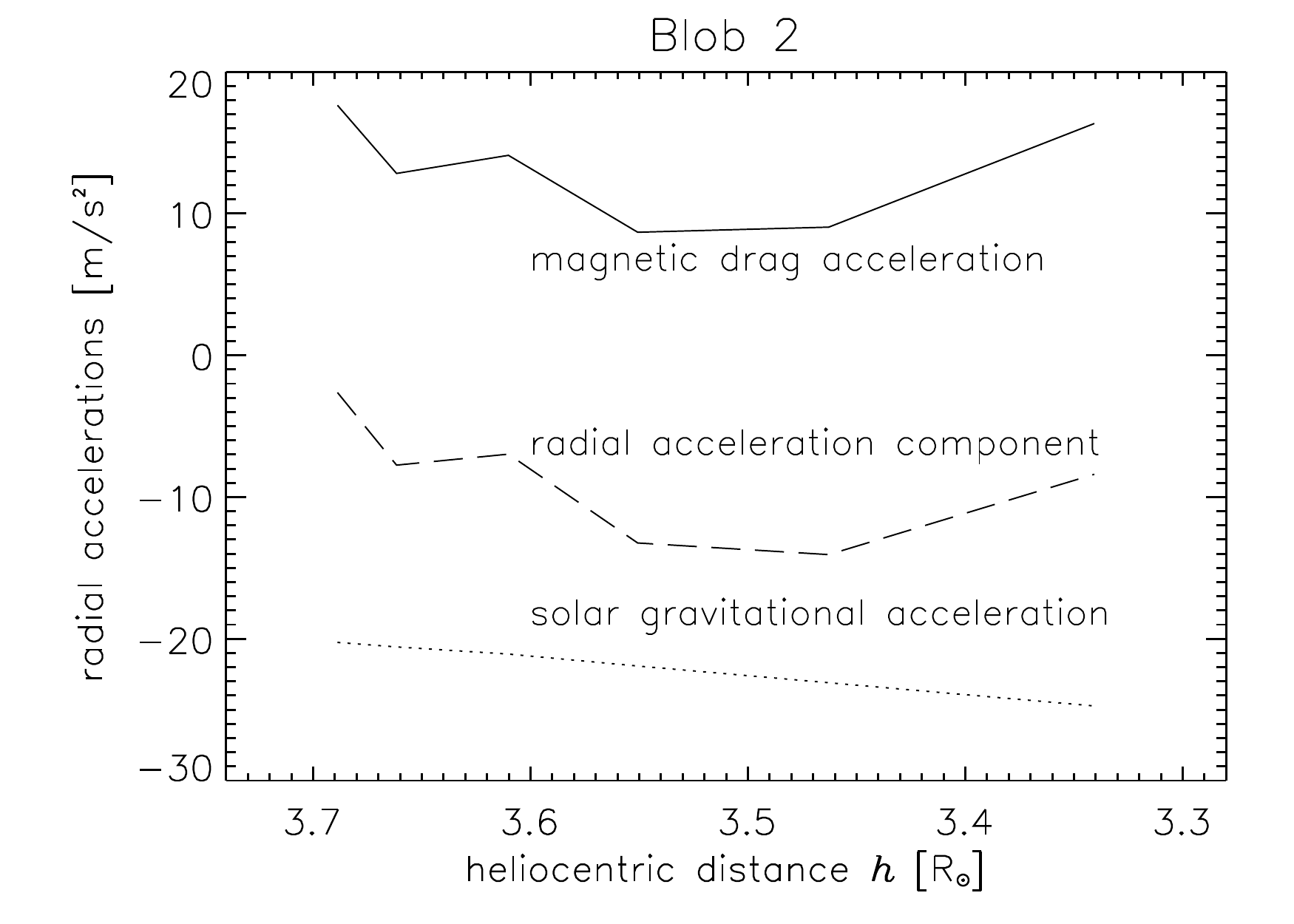}
\includegraphics[width=0.33\textwidth]{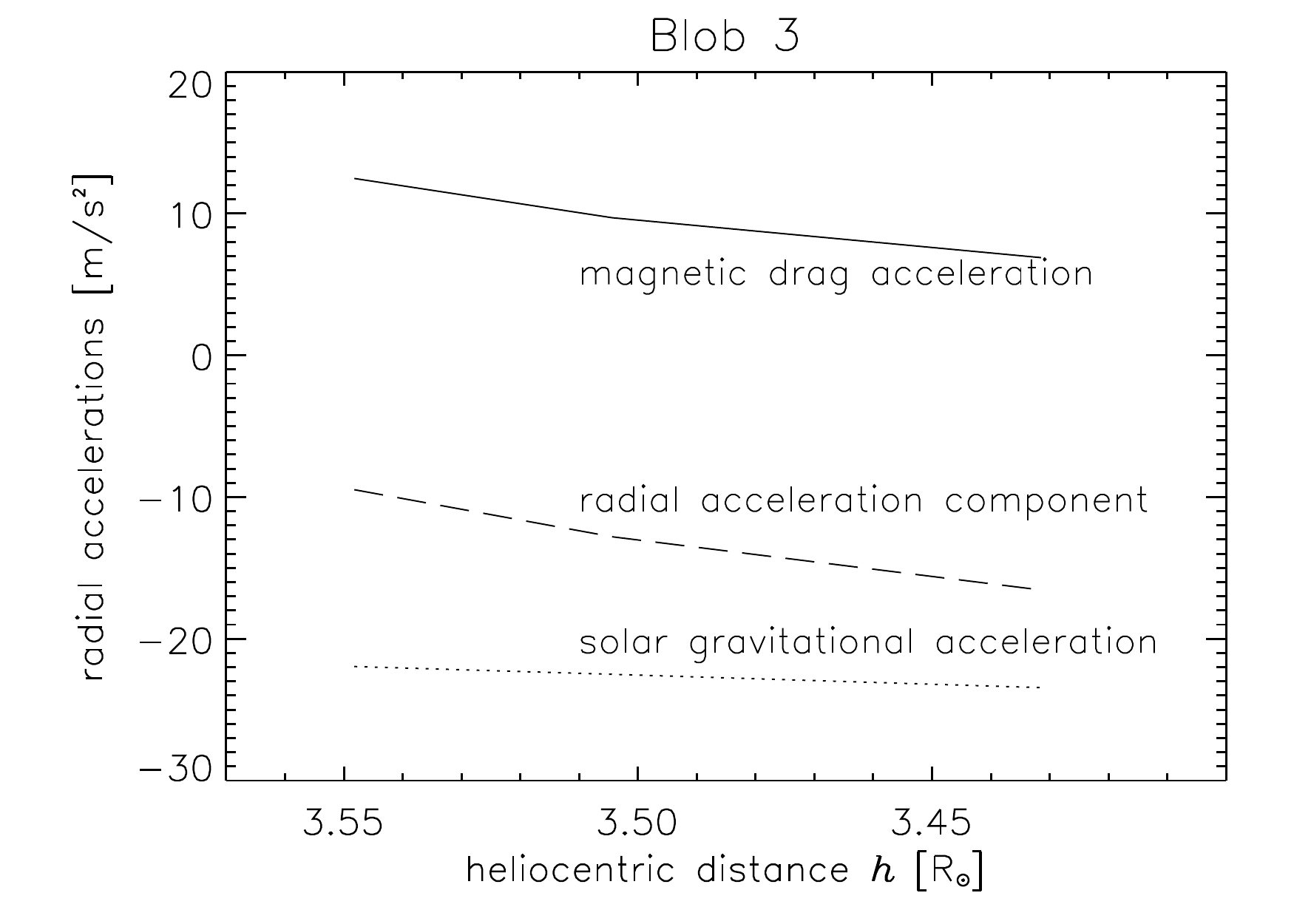}
\caption{Radial accelerations for the three blobs as a function of the heliocentric distance $h$. As a result of solar gravitational force (dotted line) and magnetic drag force (solid line) per mass unit, the blobs fall with the plotted radial acceleration component (dashed line).}
\label{arad}
\end{figure*}

\begin{figure*}
\centering
\includegraphics[width=0.33\textwidth]{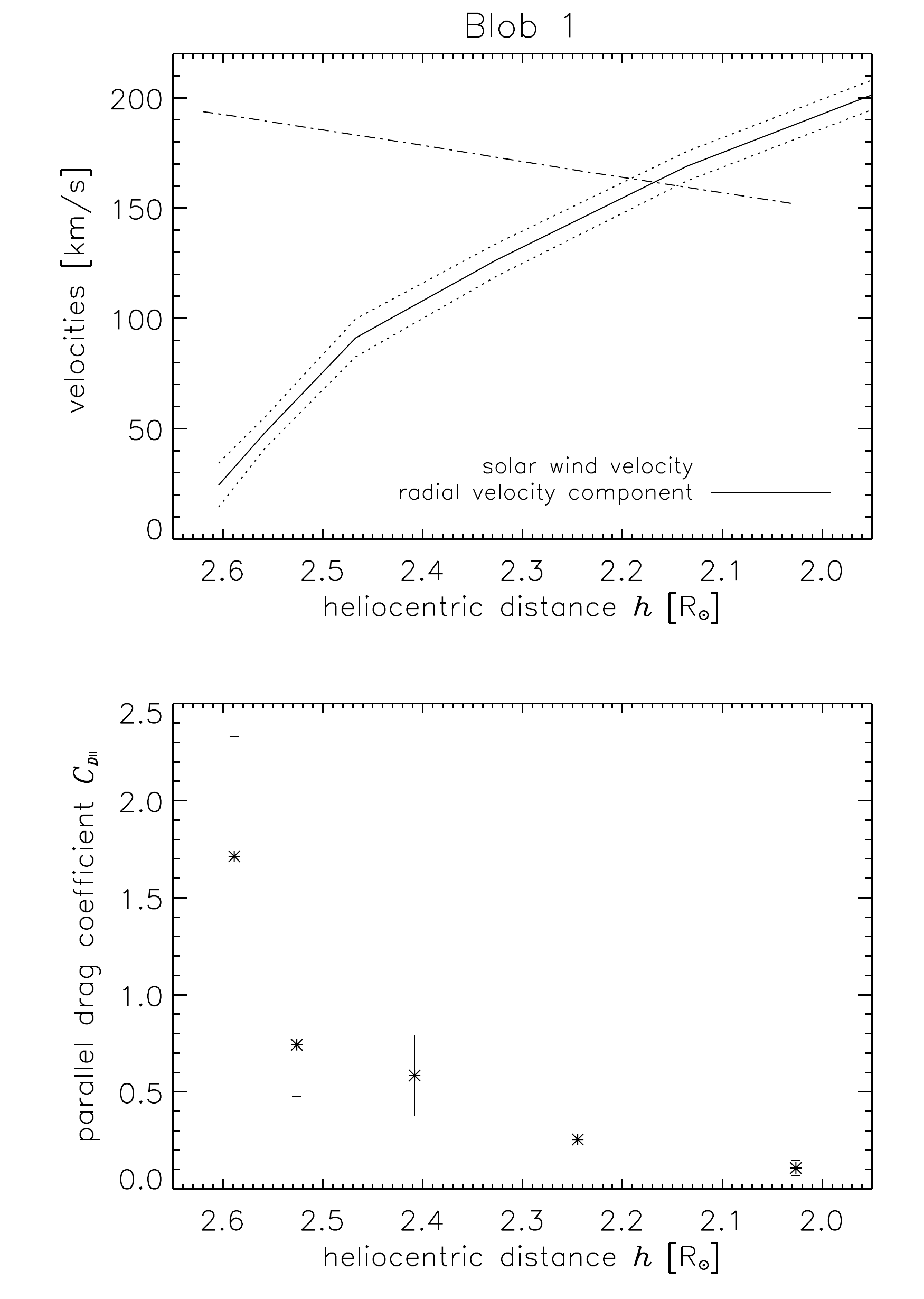}
\includegraphics[width=0.33\textwidth]{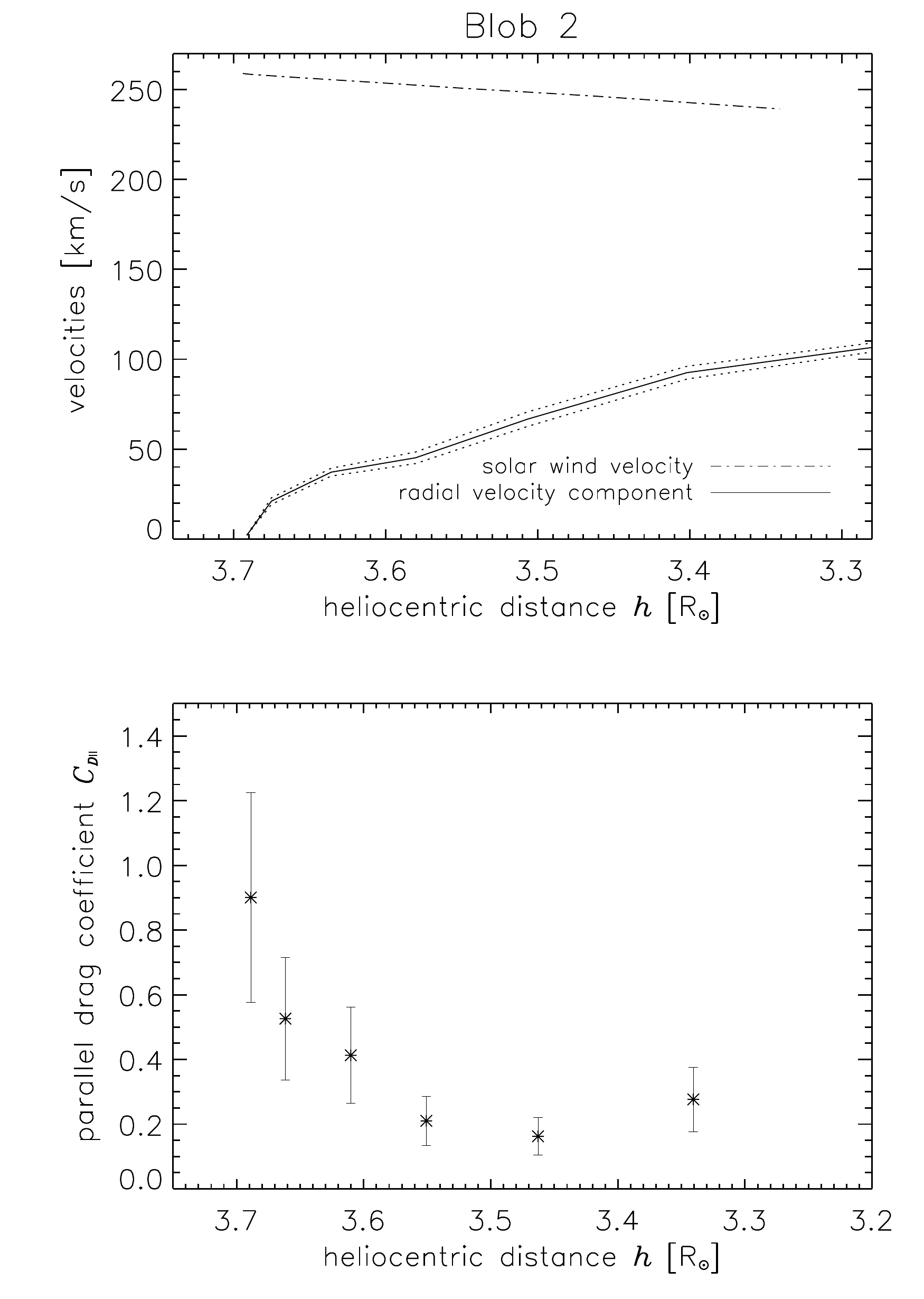}
\includegraphics[width=0.33\textwidth]{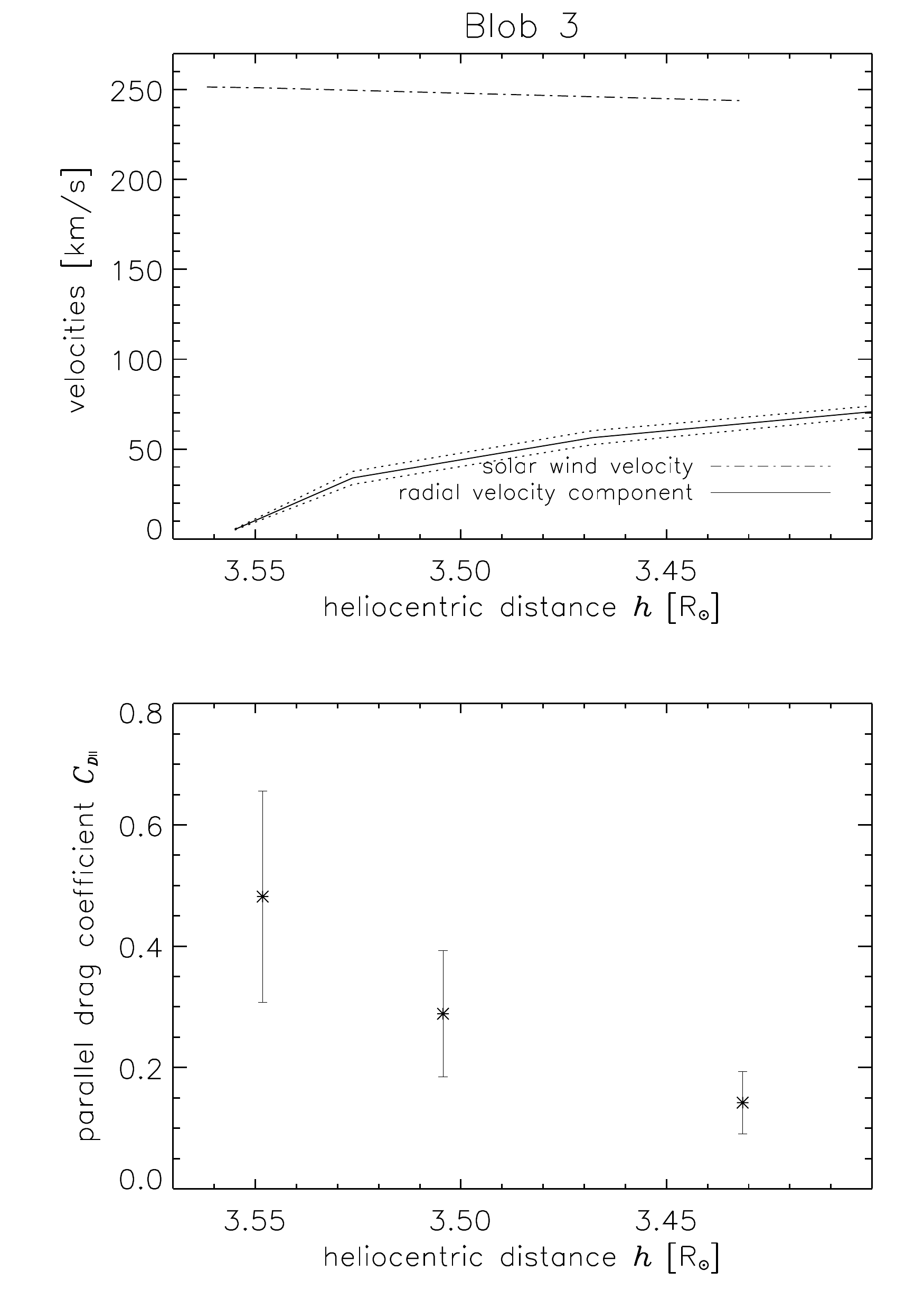}
\caption{A comparison between the modulus of $V_\parallel$ (solid line) and $V_{SW}$ (dash-dotted line), at the top, and parallel drag coefficients, at the bottom, for the three blobs at heliocentric distance $h$.}
\label{cv}
\end{figure*}

The radial component of forces acting on the blobs $F_\parallel$ is determined only by the inward gravitational force $F_g$ and the additional force due to the interaction of the blob with the solar wind, referred to as the parallel drag force $F_{D\parallel}$. The equation of motion of a blob moving in the solar wind can be written as \citep{wangsheeley2002,b2}
\begin{equation}
M_\star\frac{dV_\parallel}{dt}=M_\star g-m_p\,n_{e,cor}\,A\,C_{D\parallel}\left(V_\parallel-V_{SW}\right)\left|V_\parallel-V_{SW}\right|,
\label{cargill}
\end{equation}
where $n_{e,cor}=\langle\,\!n_{e,cor}\,\!\rangle_{2r_b}$ is the coronal density nearby the blob, calculated starting from Eq. (\ref{profile}); $g=g(h)$ is the solar gravitational acceleration; $m_p$ is the proton mass; $A=\pi\,r_b^2$ is the blob cross-sectional area; $C_{D\parallel}$ is the dimensionless parallel magnetic drag coefficient; $V_\parallel$ and $V_{SW}$ are the radial speed of the blob and of the solar wind (assumed to be positive for outflows from the Sun), respectively; and $M_\star=4\pi/3\,r_b^3\,m_p(n_{e,blob}+n_{e,cor}/2)$ is an effective mass which takes into account the real blob mass $M_b$ and the so-called virtual mass \citep{b2a}. The solar wind velocity $V_{SW}$ only has a radial component and was modeled as a function of the heliocentric distance $h$ by \citet{b3} as:
\begin{equation}
V_{SW}(h)=110+445\left(1-\frac{1}{h}\right)^{3.47}
\label{vsw}
\end{equation}
with $h$ in R$_\odot$ and $V_{SW}$ in km~s$^{-1}$. The differential term in (\ref{cargill}) is the radial acceleration component of the blob and by subtracting the $g(h)$ acceleration from it, the magnetic drag acceleration $a_D$ is derived. Finally, the parallel magnetic drag coefficient $C_{D\parallel}$ is given by
\begin{equation}
C_{D\parallel}=-\frac{4}{3}\,r_b\,a_D\left(\frac{n_{e,blob}}{n_{e,cor}}+\frac{1}{2}\right)\left[\,\left(V_\parallel-V_{SW}\right)\left|V_\parallel-V_{SW}\right|\,\right]^{-1}.
\end{equation}
Figure \ref{arad} shows the radial acceleration affecting the three blobs. We thus find that the parallel drag force acting during the falling trajectory is only a factor of 0.45~--~0.75 smaller than the solar gravitational force. Within the altitude range of 2.0~--~3.7~R$_\odot$, the solar wind velocity $V_{SW}$ given by Eq. (\ref{vsw}) changes from 150~km~s$^{-1}$ to 260~km~s$^{-1}$ and, by substituting the electron densities $n_{e,blob}$ and $n_{e,cor}$ calculated at the different blob altitudes, we obtain the parallel drag coefficient 0~$\lesssim C_{D\parallel}\lesssim$~2.5. These values are in very good agreement with what is expected by \citet{b2a} for the propagation of interplanetary CMEs; they found 0~$\lesssim$~$C_D\lesssim$~3, depending on the orientation of the flux rope and the background magnetic field (aligned or non-aligned). For each of the three blobs, in Fig. \ref{cv} (top) we show a comparison between the modulus of $V_\parallel$ and $V_{SW}$ and (bottom) the parallel drag coefficients, both as a function of heliocentric distance $h$; the $C_{D\parallel}$ profiles decrease at lower altitudes as well as in the \citet{b2a} measurements.

\subsubsection{Tangential kinematics: the Lorentz force}
\label{tan_kin}
To derive an expression for the tangential drag force $F_{D\perp}$, we start from the equation of motion of a blob crossing perpendicularly through a radial magnetic field. Similarly to \citet{b7}, and by ignoring the pressure term of the Lorentz force, one can write
\begin{equation}
m_p\,n_{e,blob}\,a_\perp\,\frac{4\pi}{3}\,r_b^3=C_{D\perp}\,\frac{V_\perp}{V_{AW}}\frac{2B^2}{\mu_0}\,2\pi\,r_b^2,
\label{H&B}
\end{equation}
where $V_\perp$ and $a_\perp$ are the tangential velocity and acceleration component, $B$ is the unknown ambient magnetic field, $\mu_0$ is the vacuum permeability, and $V_{AW}=B/\sqrt{\mu_0\,m_p\,n_{e,cor}}$ is the unknown speed of Alfv\'en waves.

In general, for a radial magnetic field $\vec{B}=B_0\,\hat{r}$, the vector operator in the expression of the magnetic tension can be reduced as
\begin{equation}
(\vec{B}\cdot\vec{\nabla})\vec{B}=B_0\frac{\partial B_0}{\partial r},
\end{equation}
and by assuming that only the magnetic tension provides the tangential (i.e. non-radial) acceleration component, a simplification of the equation of tangential motion can be given by
\begin{equation}
m_p\,n_{e,blob}\,a_\perp=\frac{B_0}{\mu_0}\frac{\partial B_0}{\partial r}\sim\frac{B_0^2}{\mu_0\,r_b}.
\label{H&B1}
\end{equation}
Given the previously derived $n_{e,blob}$ values, $r_b$ from the COR1 images and $a_\perp$ via triangulation, the radial magnetic field $B_0$ can be inferred from Eq. (\ref{H&B1}). We obtain $B_0\sim$~0.060$\pm$0.010~G at 2.5~R$_\odot$, from the blob 1 measurements, and $B_0\sim$~0.020$\pm$0.004~G at 3.5~R$_\odot$, from the blob 2 and 3 measurements.

These values are smaller by a factor of 4~--~5 than those given at the same altitudes by the \citet{b4} empirical formula, which is consistent with observations to within a factor of about 3. An estimate of the average coronal magnetic field was also provided, using the Faraday rotation technique, by \citet{b14c} from measurements on Helios in situ data, and more recently by \citet{b12a} from ground based data: their results are one order of magnitude larger in the same altitude ranges. Instead, the local coronal magnetic field value we obtained at 3.5~R$_\odot$ is comparable with that measured by \citet{b0c} at 4.3~R$_\odot$ based on the analysis of a CME-driven shock.

\begin{figure*}
\centering
\includegraphics[width=0.33\textwidth]{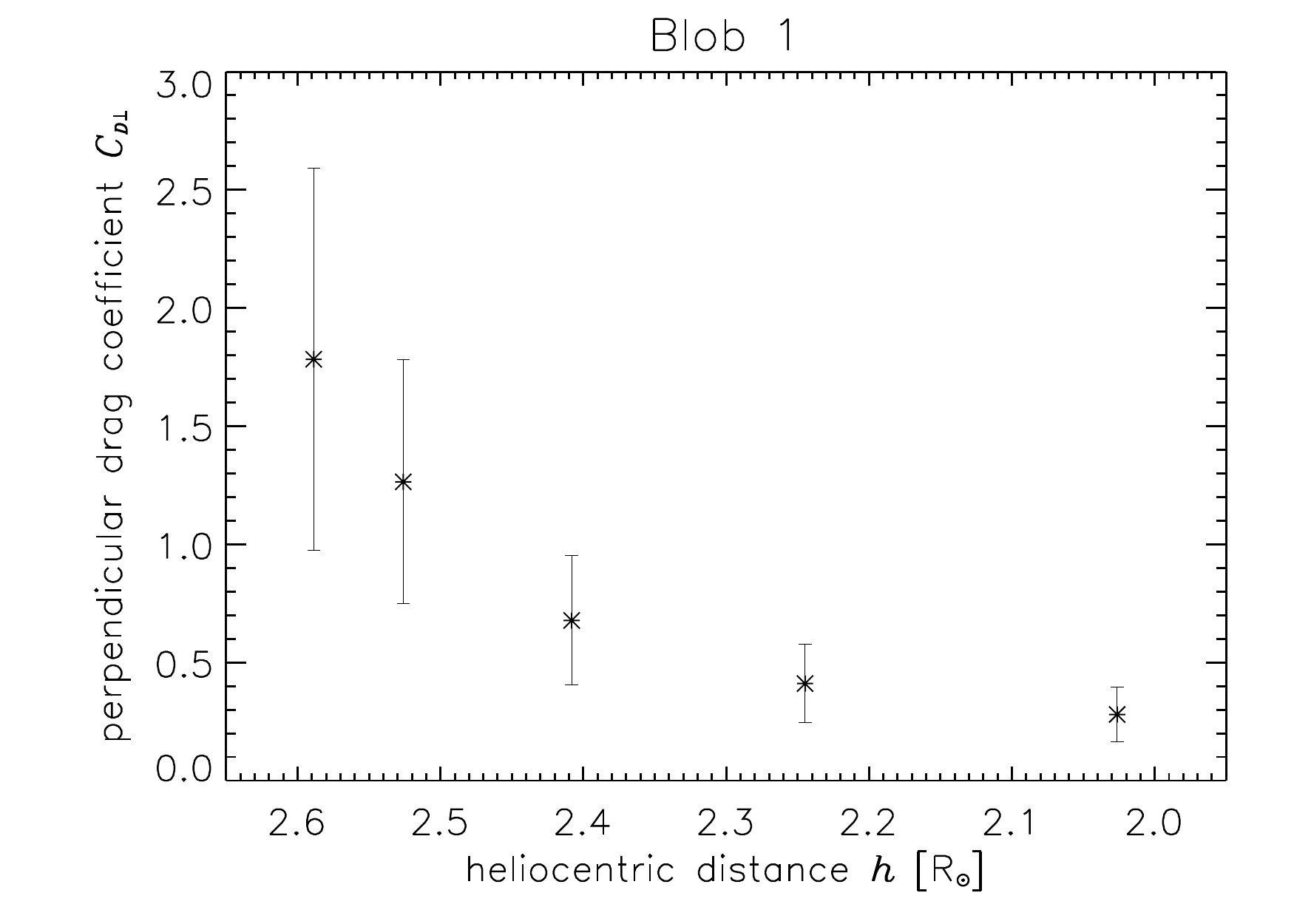}
\includegraphics[width=0.33\textwidth]{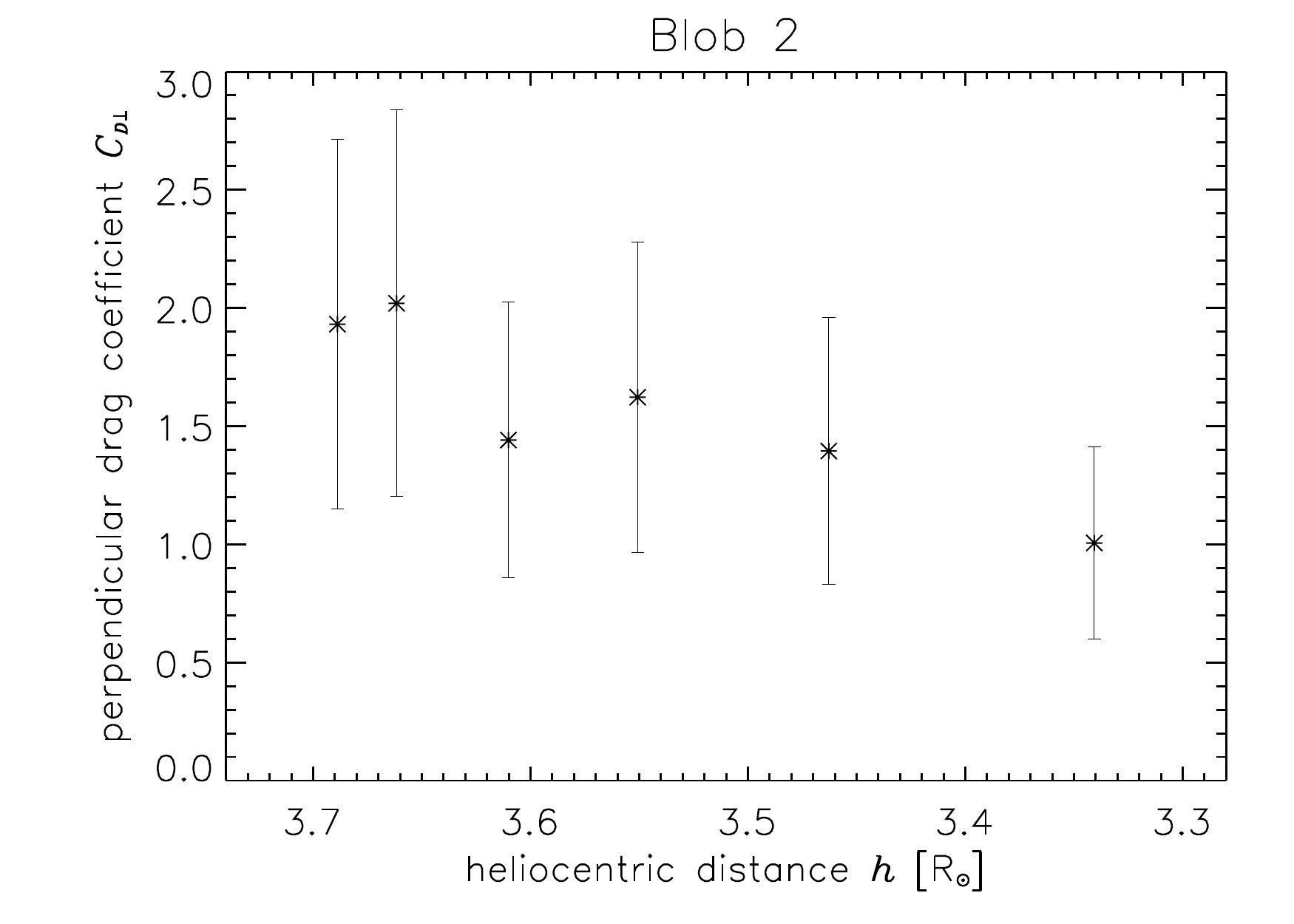}
\includegraphics[width=0.33\textwidth]{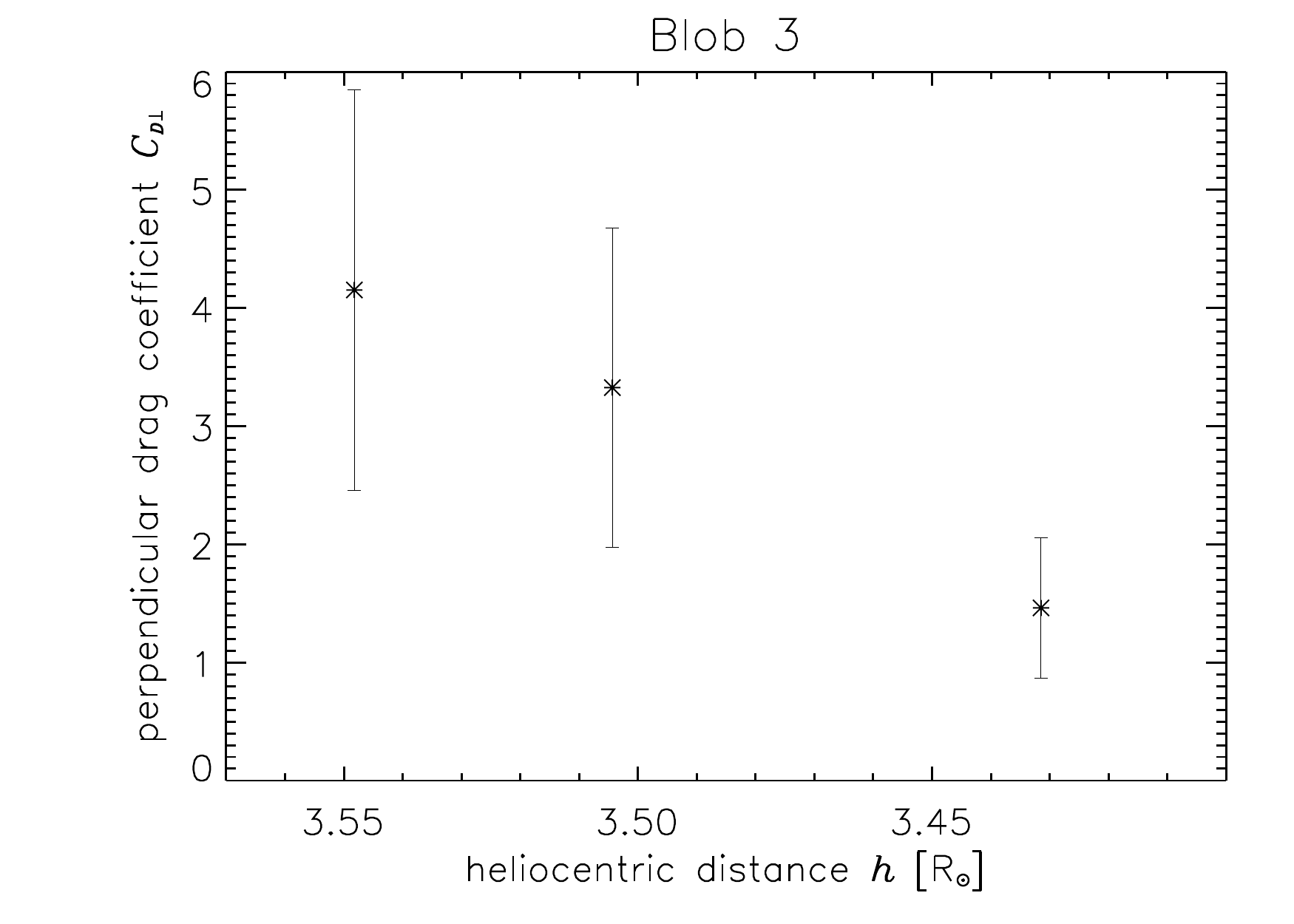}
\caption{Perpendicular magnetic drag coefficients for the three blobs at heliocentric distance $h$.}
\label{cmag}
\end{figure*}

By substituting the expression for $B$ from Eq. (\ref{H&B1}) into Eq. (\ref{H&B}), the perpendicular magnetic drag coefficient $C_{D\perp}$ is given by
\begin{equation}
C_{D\perp}=\frac{1}{3\,V_\perp}\sqrt{\frac{n_{e,blob}r_b\,a_\perp}{n_{e,cor}}}=\frac{\sqrt{X\,r_b\,a_\perp}}{3\,V_\perp}.
\end{equation}
This equation allows us to easily estimate $C_{D\perp}$ once $V_\perp$, $a_\perp$ and the density ratio $X=n_{e,blob}/n_{e,cor}$ are known. In the end, the perpendicular magnetic drag coefficients turn out to be 0.2~$\lesssim$~$C_{D\perp}\lesssim$~5 (see Figure \ref{cmag}). Hence $C_{D\parallel}\sim C_{D\perp}$.

\subsubsection{Magnetic pressure in the blobs}

As the COR1 images show, most of the chromospheric plasma assembles as blobs, which fade away during their falling motion towards the Sun. This is probably due to an imbalance between internal and external kinetic and magnetic pressure. If no such imbalance exists, the magnetic field trapped in the chromospheric plasma blobs can be estimated by assuming that internal and external total pressure are balanced. Hence, in order to explain the persistence of the blobs as coherent structures, in what follows we verify that, by assuming reliable coronal temperatures and imposing pressure balance, our results also correspond to reliable values of the coronal magnetic field. In the hypothesis of pressure balance we can write that
\begin{equation}
2\,n_{e,blob}\,k_B\,T_b+\frac{B_b^2}{2\mu_0}=2\,n_{e,cor}\,k_B\,T_e+\frac{B_0^2}{2\mu_0},
\end{equation}
where $k_B$ is the Boltzmann constant; $T_b$ and $B_b$ are temperature and magnetic field inside the blob, respectively; and $T_e$ is the temperature of the surrounding corona. \citet{b3} modeled the coronal hole temperature as a function of the heliocentric distance $h$ and obtained
\begin{equation}
T_{e,cor}(h)=10^6\times\left(0.35\,h^{1.1}+1.9\,h^{-6.6}\right)^{-1}
\label{tcor}
\end{equation}
from which $T_e$ can be calculated; in Eq. (\ref{tcor}), $h$ is in units of R$_\odot$ and $T_{e,cor}$ is in Kelvin. As already assumed in Sect.~\ref{bdc} for calculating the electron density of the blobs, $T_b=$ 20000~K and we achieve a value of $B_b$ about equal to 0.082$\pm$0.020~G for blob 1 and 0.040$\pm$0.007 G for blobs 2 and 3.

\section{Discussion and conclusions}
\label{conclusions}

In this work we provided the first estimates of the magnetic drag force acting on small-scale ($\sim0.01$~R$_\odot$) plasma blobs falling in the intermediate corona in the heliocentric distance range 2.0~--~3.7~R$_\odot$. To this end, we recovered the 3-D kinematics of the plasma blobs ejected into the corona during the huge flare of June 7, 2011; we performed this analysis using STEREO/COR1 coronagraphic data with a proven triangulation technique \citep{b9}. The motion of ejected blobs falling back to the Sun differs from the simple ballistic motion due to solar gravity alone: we ascribed this difference to the presence of magnetic drag acting on these blobs. The blobs were ejected with significant velocity components both in the radial and tangential directions. Hence, in order to estimate the drag force, we made a distinction between radial and tangential drag forces and we ascribed the latter to the draping of the radial coronal field crossed by the plasma blobs. Our analysis led us to conclude that the magnetic drag forces acting on small-scale blobs falling in the intermediate corona are considerable, being only a factor 0.45~--~0.75 smaller than the gravitational force.

The plasma blobs studied here are very likely not representative of the kinematics of CMEs as a whole. Nevertheless the possibility that the plasma physics processes at play during the propagation of small-scale blobs leading to the effective drag reported here are similar to those playing a role during the propagation of large-scale CMEs cannot be ruled out in principle. It is widely accepted that CMEs are mainly driven and accelerated in the lower corona by the Lorentz forces related to magnetic field lines involved in the eruption \citep{b17c}. It is also well known that in the interplanetary medium CMEs and solar wind are dynamically coupled by the occurrence of a magnetic drag, which is probably the resulting large scale manifestation of magnetohydrodynamics waves being produced by the CME-wind interaction carrying away momentum and energy of the CME \citep{b3,b2a}; this interaction makes the drag force dominant and causes the CME velocities to converge to the solar wind velocities \citep{b4b}. Nevertheless, the balance of forces governing the CME evolution during both the early acceleration and the later propagation phases is unclear. For instance, \citet[][]{foullon2011} recently demonstrated that plasma vortices may form in the lower corona at the flanks of CMEs (because of Kelvin-Helmholtz instability) and pointed out that an important direct consequence of these vortices ``is their effect on the total drag force, which affects the CME kinematics and hence its geoeffectiveness''. This was the first observational evidence of the possible role played by magnetic drag forces in the early evolution of CMEs. Nevertheless, at present there are no published works providing estimates of these forces in the lower corona, hence their possible role in the early evolution of CMEs is in general unknown or possibly underestimated, and numerical models usually simply assume these forces to be negligible with respect to the Lorentz and gravitational forces. The quantitative results described here show that, in the intermediate corona, drag forces affecting small-scale blobs are significant and cannot be neglected when describing the dynamics of these blobs. If these forces are similar to those affecting the early propagation of CMEs, our results suggest that the magnetic drag force should be also considered in any CME initiation model.

In the course of explaining the kinematics of the blobs, we also provided new techniques for analysing COR1 data in order to 1) estimate the electron density in chromospheric plasma blobs raised into in the corona, 2) separate in COR1 data the white-light intensity due to Thomson scattering and to H$\alpha$ emission, and 3) estimate the degree of polarization of H$\alpha$ emission. Very recently, \citet[][]{williams2013} published measurements of column densities for electrons and hydrogen atoms in blobs ejected as a consequence of the same eruption reported here. Densities reported by these authors obtained with SDO/AIA data analysis are much larger than those reported here (in the order of $\sim$~1~--~3~$\times10^{10}$~cm$^{-3}$); nevertheless, these densities are relative to plasma observed by AIA telescopes on disc, hence at much lower altitudes than blobs reported here in the altitude range 2.0~--~3.7~R$_\odot$, making a direct comparison meaningless. Other plasma blobs ejected during the same eruption were also the subject of a very recent work by \citet[][]{b14g}, where the kinetic energy being dissipated as the blobs collided with the chromosphere was measured. Considering again the much larger density of blobs they analysed \citep[analogous to those studied by][]{williams2013} with respect to those reported here, the authors concluded that the local magnetic pressure was negligible over the ram and thermal pressures, hence the trajectories of the blobs were close to simple free falls. This is different from the present case, where the magnetic pressure is not negligible, as we discussed.

The electron densities in the blobs were derived in the present work from observed white-light intensity after the removal of H$\alpha$ contribution. This last component was estimated by assuming the expression given by \citet{b0b} for the ratio between the H$\alpha$ radiation and emission by Thomson scattering in chromospheric prominence plasma. Results show that the H$\alpha$ contribution to the total white-light intensity observed at the location of the blobs is between 95 and 98~\%, making the subtraction of this contribution crucial for a correct estimate of plasma densities. In addition, taking advantage of COR1 polarized brightness measurements, we demonstrated that COR1 data can also be employed to estimate the amount of H$\alpha$ polarized emission, which is in the order of a few percent of H$\alpha$ total emission. This is the first time that the amount of H$\alpha$ polarized emission has been estimated in the intermediate corona between 2.0 and 3.7~R$_\odot$. Measurements of this quantity are known to be very important because they are able to provide unique information on the magnetic field embedded in the prominence, via the so-called Hanle effect \citep[see e.g.][and references therein]{b0a}. Hence, our results not only demonstrate the capability for future determinations of H$\alpha$ unpolarized and polarized emission with COR1 data, but also demonstrate that this polarized component is significant, thus opening the possibility for future measurements of magnetic fields in erupting prominences in the intermediate corona with coronagraphic images.

\begin{acknowledgements}
This work was partly supported by the Agenzia Spaziale Italiana through the contracts ASI/INAF I/023/09/0, I/043/10/0, I/013/12/0 and by the European Commissions Seventh Framework Programme (FP7/2007-2013) under the grant agreement SWIFF (project no. 263340). The authors thank P. Heinzel for very important comments on the analysis of H$\alpha$ radiation.  The SECCHI data are produced by an international consortium of the NRL, LMSAL and NASA GSFC (USA), RAL and University of Bham (UK), MPS (Germany), CSL (Belgium), and IOTA and
IAS (France).
\end{acknowledgements}

\end{document}